\documentclass[twocolumn,trackchanges]{aastex631}

\begin{document}

\title{\textit{Kepler} and \textit{TESS} Observations of PG~1159-035}

\author{Gabriela Oliveira da Rosa}
\affiliation{Instituto de F\'{\i}sica, Universidade Federal do Rio Grande do Sul \\ 91501-970 Porto Alegre, RS, Brazil\\}

\author{ S. O. Kepler}
\affiliation{Instituto de F\'{\i}sica, Universidade Federal do Rio Grande do Sul \\ 91501-970 Porto Alegre, RS, Brazil\\}

\author{Alejandro H. C\'orsico}
\affiliation{Facultad de Ciencias Astron\'omicas y Geof\'{\i}sicas, Universidad Nacional de La Plata \\ Paseo del Bosque s/n, 1900, Argentina\\}
\affiliation{IALP - CONICET, La Plata, Argentina\\}

\author{J. E. S. Costa}
\affiliation{Instituto de F\'{\i}sica, Universidade Federal do Rio Grande do Sul \\ 91501-970 Porto Alegre, RS, Brazil\\}

\author{J. J. Hermes}
\affiliation{Department of Astronomy \& Institute for Astrophysical Research, Boston University \\ 725 Commonwealth Ave., Boston, MA 02215, USA\\}

\author{S. D. Kawaler}
\affiliation{Department of Physics and Astronomy, Iowa State University\\
Ames, IA 50011, USA \\}

\author{Keaton J. Bell}
\affiliation{DIRAC Institute, Department of Astronomy, University of Washington\\ Seattle, WA-98195, USA \\}

\author{M. H. Montgomery} 
\affiliation{Department of Astronomy, University of Texas at Austin\\ Austin, TX-78712, USA \\}
\affiliation{McDonald Observatory\\ Fort Davis, TX-79734, USA \\}

\author{J. L. Provencal} 
\affiliation{Department of Physics and Astronomy Newark, University of Delaware \\ DE 19716, USA \\}
\affiliation{Delaware Asteroseismic Research Center, Mt. Cuba Observatory\\ Greenville, DE 19807, USA \\}

\author{D. E. Winget} 
\affiliation{Department of Astronomy, University of Texas at Austin\\ Austin, TX-78712, USA \\}
\affiliation{McDonald Observatory\\ Fort Davis, TX-79734, USA \\}

\author{G. Handler} 
\affiliation{Nicolaus Copernicus Astronomical Center, Polish Academy of Sciences\\ Bartycka 18, 00–716 Warsaw, Poland \\}

\author{Bart Dunlap}
\affiliation{Department of Astronomy, University of Texas at Austin\\ Austin, TX-78712, USA \\}
\affiliation{McDonald Observatory\\ Fort Davis, TX-79734, USA \\}

\author{J. C. Clemens}
\affiliation{Physics and Astronomy Department, University of North Carolina at Chapel Hill\\ Chapel Hill, NC 27599 \\}

\author{Murat Uzundag}
\affiliation{Instituto de F\'isica y Astronom\'ia, Universidad de Valpara\'iso \\ Av. Gran Breta\~na 1111, Playa Ancha, Valpara\'iso 2360102, Chile \\}
\affiliation{European Southern Observatory \\ Alonso de Cordova 3107, Santiago, Chile \\}



\begin{abstract}
PG~1159-035 is the prototype of the DOV hot pre-white dwarf pulsators.  It  was observed during the Kepler satellite {\it K2} mission for 69 days in 59~s cadence mode and by the {\it TESS} satellite for 25 days in 20~s cadence mode. We present a detailed asteroseismic analysis of those data.  We identify a total of 107 frequencies  
representing 32 $\ell=1$ modes, 27 frequencies representing 12 $\ell=2$ modes, and 8 combination frequencies.  The combination frequencies and the modes with very high {\it k} values represent new detections. The multiplet structure reveals an average splitting of $4.0\pm0.4\ \mu$Hz for $\ell$=1 and $6.8\pm0.2\ \mu$Hz for $\ell=2$, indicating
a rotation period of $1.4\pm0.1$ days in the region of period formation. In the Fourier transform of the light curve, we find a significant peak at $8.904\pm0.003\ \mu$Hz suggesting a surface rotation period of $1.299\pm0.002$ days.
We also present evidence that the observed periods change on timescales shorter than those predicted by current evolutionary models. Our asteroseismic analysis finds an average period spacing for $\ell=1$ of $21.28\pm0.02$~s. The $\ell=2$ modes have a mean spacing of $12.97\pm0.4$~s.  We performed a detailed asteroseismic fit by comparing the observed periods with those of evolutionary models. The best fit model has $T_\mathrm{eff}=129\,600\pm 11\,100$~K, $M_*=0.565\pm 0.024 M_{\odot}$, and $\log g=7.41^{+0.38}_{-0.54}$, within the uncertainties of the spectroscopic determinations. We argue for future improvements in the current models,  e.g., on the overshooting in the He-burning stage, as the best-fit model does not predict excitation for all the pulsations detected in PG~1159-035.
\end{abstract}


\section{Introduction}
\label{section1}
\par White dwarf (WD) stars are the evolutionary end point of all stars born with masses up to $\simeq 10.5~M_\odot$, which correspond to more than 98\% of all stars \citep[e.g.][]{Lauffer18}. The effective temperature of WDs  ranges from $T_\mathrm{eff}\simeq 200\,000~$~K to around $4500$~K, and masses from $\simeq 0.15~\mathrm{to}~\simeq 1.36~M_\odot$.
\par PG~1159-035 is the prototype of the hot WD spectroscopic class called PG~1159, as well as the GW Vir class of pulsating variable stars (PG~1159-035 = GW Vir = DOV) \citep{1979wdvd.coll..377M, 2019A&ARv..27....7C}. The PG~1159 spectroscopic class is characterized by a strong H deficiency and high-excitation He~II, C~IV, O~VI and N~V lines \citep[e.g.][]{Werner89,Sowicka21, Werner22}. These are among the hottest pulsating stars known.
\par The pulsation modes observed in WDs are nonradial $g$ (gravity) modes. Gravity acts as the restoring force on the displaced portions of mass, moving it mainly horizontally. These pulsations cause different temperature zones that oscillate at eigenfrequencies, restricted by the spherical symmetry of the star.
\par In asteroseismology, we describe a pulsation mode using a spherical harmonic basis with three integer quantum numbers: $k$, $\ell$ and $m$. The number $k$ is called the radial index and is the number of radial nodes, related to how ``deep'' a mode is located in the star. 
The larger the radial index of a mode, the more superficial is its main region of period formation. The number $\ell$ is called the spherical harmonic index and is related to the number of latitudinal hot and cold zones. Finally, the number $m$ is called the azimuthal index, and its absolute value is related to the arrangement of those zones on the stellar surface. The number $m$ assumes integer values from $-\ell$ to $+\ell$.  Rotation of the star breaks the degeneracy of the pulsation modes with same $k$ and $\ell$ but different $m$, causing the modes to split into $2\ell+1$ components in the Fourier Transform (FT) of its light curve.

\par Due to geometrical cancellation, we expect to observe predominantly modes with $\ell=1$ and $\ell=2$ in WDs \citep{1982ApJ...259..219R}. These modes should produce triplets and quintuplets in Fourier Transforms (FT) of light curves of rotating WDs. This expectation is supported by the work of \citet{Stahn05}.  The authors make use of the wavelength dependent flux variations, or chromatic amplitudes, for modes with different $\ell$.  They extracted the chromatic amplitudes from 20 orbits of $\it{HST-STIS}$ time resolved spectra of PG~1159-035 between 1100~\AA\ and 1750~\AA. Comparing the results to models, they concluded that the most prominent pulsation mode at 516~s matches $\ell=1$ or $\ell=2$ modes only.

\section{Previous datasets}
\label{section2}
PG~1159-035 has been observed by different ground-based telescopes since 1979 (Table~\ref{obs_data}). The ground-based data consist primarily of photometric observations obtained with CCDs and photomultiplier tubes.  The Whole Earth Telescope (WET) runs in 1989, 1993, and 2002 were multi-site international campaigns dedicated to achieving 24~h coverage \citep{Winget91}.  In 2016 and 2021,  this important star was continuously observed by space-based telescopes, enabling unprecedented quality data. Table~\ref{obs_data} is a journal of the main observational campaigns since 1983. This table shows that, although the previous campaigns have comparable --- or even longer --- total lengths, the {\it K2} data~(2016) is by far the one with the most dense observations, followed by the {\it TESS} data~(2021).

\begin{table}[ht]
    \centering
    \begin{tabular}{|c|c|c|c|c|} \hline 
        Year & Telescopes & Length & On star & Spectral  \\ \newline
            & & (days) & (days) & resolution ($\mu$Hz) \\ \hline \hline \newline
        1983 & McDonald, SAAO$^a$ & 96.0 & 2.7 & 0.12 \\ \newline 
        1985 & McDonald, SAAO$^a$ & 64.6 & 2.0 & 0.18 \\ \newline 
        1989 & WET$^b$ & 12.1 & 9.5 & 0.96 \\ \newline 
        1993 & WET$^c$ & 16.9 & 14.4 & 0.68 \\ \newline 
        2002 & WET$^d$ & 14.8 & 4.8 & 0.78 \\ \newline 
        2016 & {\it Kepler} & 69.1 & 54.5 & 0.17 \\ \newline 
        2021 & {\it TESS} &24.9 & 22.0 & 0.46 \\ \hline
    \end{tabular}
\footnotesize{$^a$~\citet{Winget1985};
$^b$~\citet{Winget91}; $^c$~\citet{Bruvold1993}; \\$^d$~\citet{Costa2003}}
    \caption{Main observational campaigns of PG~1159-035 between 1983 and 2021.}
    \label{obs_data}
\end{table}
    
\par
\begin{figure*}[t]
    \centering
    \includegraphics[width=1\textwidth,trim=2.8cm 1.5cm 2.8cm 2.5cm]{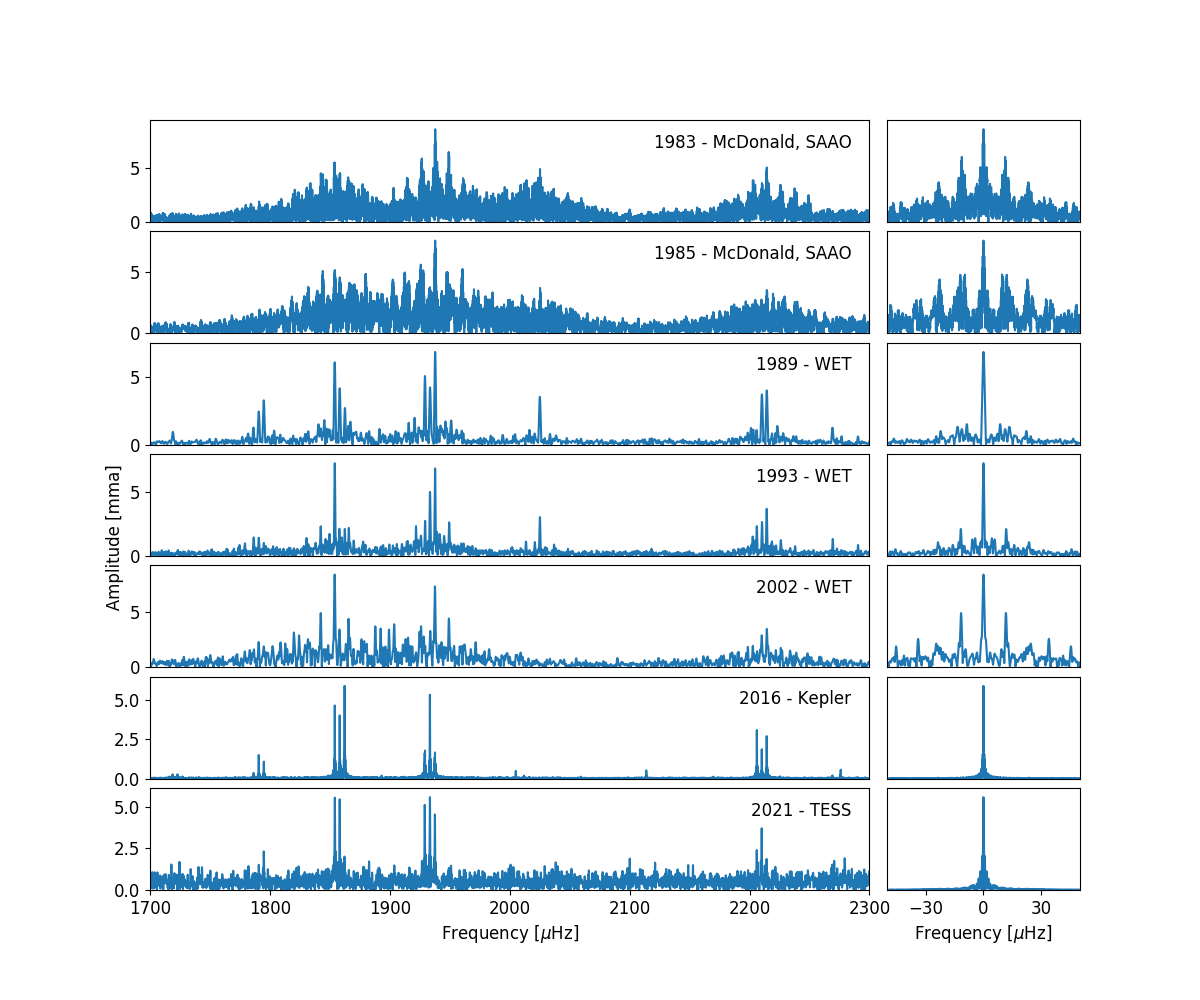}
    \caption{Fourier transform of PG~1159-035 of the years of 1983, 1985, 1989, 1993, 2002, 2016 and 2021, from 1700 to 2300 $\mu \mathrm{Hz}$. Their respective spectral windows are shown on the right side. Frequencies are in $\mu \mathrm{Hz}$ and amplitude units use 1\,mma = 1/1.086 mmag  = 1\,ppt = 0.1\%, see e.g. \citet{2016IBVS.6184....1B}.} 
    \label{FT+SW}
\end{figure*}

 Figure~\ref{FT+SW} shows the Fourier transform (FT) for each annual observation of PG~1159-035, in the range of the higher amplitude peaks (1700-2300~$\mu$Hz, or roughly $435-590$\,s), and their respective spectral window on the right side.

As shown in this figure, the {\it K2} data spectral window has the sharpest structure, allowing the triplets to appear more clearly in the FT. The {\it TESS} data spectral window also has very little structure, but the {\it TESS} data signal-to-noise ratio is limited by the small size of the telescope and the redder bandpass.  PG~1159-035 is very blue ($T_\mathrm{eff}\approx 140,000$K) and faint ($m_{v}=15.04$).  The higher noise found in the {\it TESS} data hinders the detection of the numerous low amplitude frequencies found in the {\it K2} data.

\section{{\it{K2}} and {\it{TESS}} Observational Data}
\label{section3}
\par After a failure of the second reaction wheel controlling the pointing of the {\it Kepler} spacecraft, observations along the ecliptic plane were enabled by the {\it K2} mission \citep{Howell14}. {\it K2} observed PG~1159-035 (EPIC 201214472) between July and September 2016 during Campaign 10.
We downloaded the target pixel files (TPFs) in short cadence (58.85s) from the Barbara A. Mikulski Archive for Space Telescopes (MAST), and used the {\em Lightkurve} package \citep{lk18} to extract photometry from the TPFs. As {\it K2} suffers a $\sim 6.5$~hr thruster firing to compensate for the solar pressure variation for fine pointing, we subsequently used the KEPSFF routine \citep{2014PASP..126..948V} to correct the systematic photometric variation that is induced by the low-frequency motion of the target on the CCD module. A series of apertures of different pixel sizes were tested on the TPF to optimize the photometry. We finally chose a fixed 30-pixel aperture to extract our light curve.
After extracting the photometry, we fit a third-order polynomial and sigma clipped the light curve to 4.5$\sigma$ in order to detrend the light curve and to clip the outliers. We also subtracted the known electronic spurious frequencies and their harmonics \citep{KDH16}. The {\it K2} data starts at Barycentric Julian Dates in Barycentric Dynamical Time BJD\_TDB=2457582.5799677 and extends 69.14 days, with 58.8~s cadence.
\par The {\it TESS} data were collected in 2021 December during Sector 46 with the spacecraft's fastest 20~s cadence. The data were downloaded from the MAST Portal and used {\em PDCSAP} simple aperture fluxes, after removal of 5$\sigma$ outliers.  PG~1159-035 is TIC 35062562 in the {\it TESS} Target Input Catalog.

\section{Detection of pulsation periods}
\label{section4}
\par We used the Period04 Fourier analysis software \citep{2004IAUS..224..786L} to detect the pulsation frequencies and subtract their respective sinusoids from the light curves of {\it K2} (pre-whitening). We estimated the false-alarm-probability (fap) of 1/1000 by  randomizing the input 1000 times, as the data consists of multiple coherent frequencies. 
\par After the subtraction of each peak found in the Fourier Transform (FT) directly from the light curve, we calculated the detection limit of the residual light curve. We repeated this process until the highest amplitude peak had a false alarm probability larger than $1/1000$.

\begin{figure}
    \centering
    \includegraphics[width = 0.5\textwidth, trim= 0.5cm 0 0.cm 0]{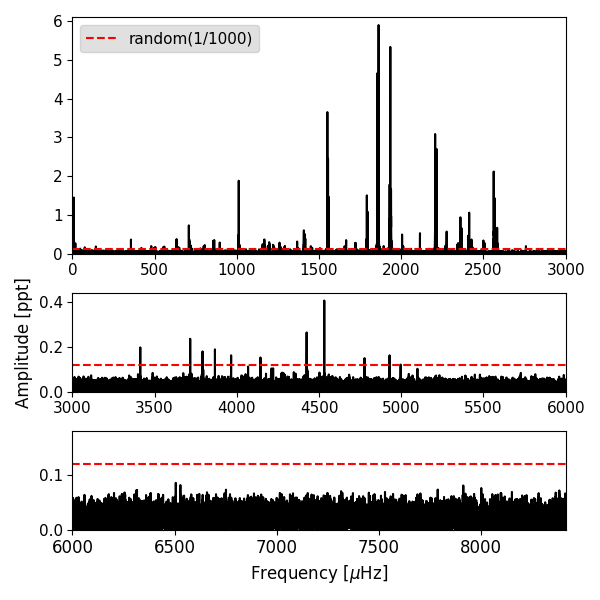}
    \caption{Fourier Transform of the original {\it K2} light curve of PG~1159-035. The red dashed line indicates the false alarm probability fap(1/1000)=0.119~mma we obtained randomizing the input times 1000 times. Frequencies are in $\mu \mathrm{Hz}$ and amplitudes in $ppt$}.
    \label{baluev}
\end{figure}

\begin{figure}
    \centering
    \includegraphics[width = 0.5\textwidth, trim= 0.8cm 0 0.cm 0]{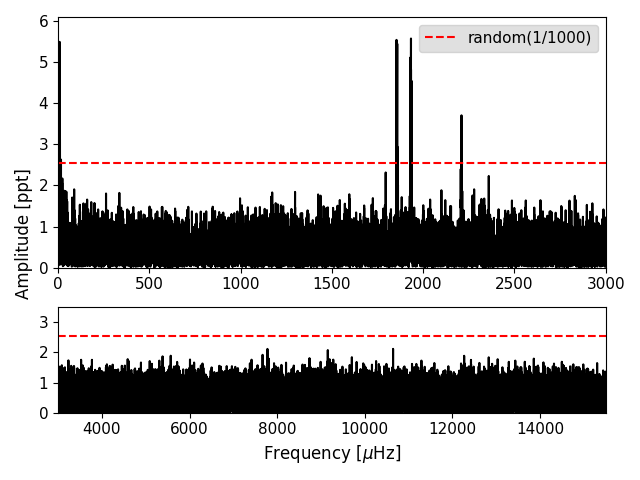}
    \caption{Fourier Transform of the original {\it TESS} light curve of PG~1159-035. The red dashed line indicates the false alarm probability fap(1/1000)=~2.537~mma we obtained randomizing the input times 1000 times.}
    \label{baluev1}
\end{figure}

\par Tables \ref{tab:l=1} and  \ref{tab:l=2} list the $\ell=1$ and $\ell=2$ frequencies detected in the {\it K2}, and Table \ref{tab:tess} in the {\it TESS} data sets. The values for the periods given in the tables were determined by non-linear simultaneous multisinusoidal least-squares fit and their uncertainties by Monte Carlo simulation, as in \citet{Costa08}.

\section{Mode coherence}
\label{section5}
\par
\begin{figure}[ht]
    \centering
    \includegraphics[width=0.5\textwidth,trim= 1.5cm 0cm 0.5cm 1cm]{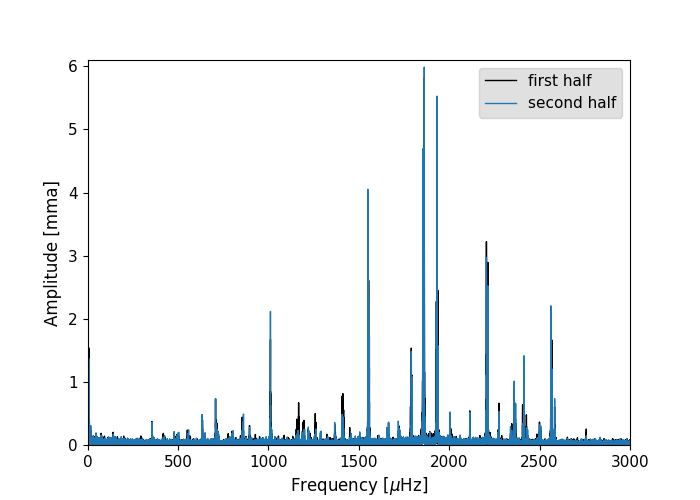}
    \caption{Fourier transform of first half and second half of the K2 data, showing amplitude changes during the 69~d observations, especially for peaks between 1100 and 1400~$\mu$Hz.}
    \label{fig:halves}
\end{figure}

 Asteroseismic analysis is based on the detected mode properties. The basic underlying assumption is that the frequencies and amplitudes of the pulsations modes are stable on a much longer baseline than that of the observations. 
Changes in stellar structure do affect the amplitude and frequency of pulsation modes.  
The Fourier transforms of PG~1159-035 from different epochs show different mode frequencies and amplitudes 
(Figures~\ref{FT+SW} and \ref{fig:halves}), 
indicating that its pulsation modes are not strictly coherent on time scales of months or years. 

The intrinsic width of a mode in the power spectra is inversely related with its lifetime. A coherent mode appears in the power spectra as a single peak with a width dictated by
the length of the observations.  Such a peak has a lifetime considerably longer than the time span of the data set. On the other hand, if the observational campaign is lengthy enough to observe them, an incoherent mode with a short lifetime appears in the power spectra as a multitude of closely spaced peaks \citep[e.g.][]{Basu17}.
\par Motivated by the results presented in \citet{2017ApJS..232...23H} about the dichotomy of mode widths for ZZ~Ceti  (DAV\footnote{Cool pulsating white dwarfs with hydrogen atmosphere.}) stars, we fitted  Lorentzian envelopes by least-squares to every set of peaks detected in the {\it K2} data power spectrum, as in \citet{2015ApJ...809...14B}.  We used the highest-amplitude peak within each set of peaks as an initial guess for the central frequency and Lorentzian height. For the half-width at half-maximum (HWHM), we take twice the frequency resolution as an initial guess. 

\begin{figure*}
    \centering
    \includegraphics[width=\textwidth, trim= 0.6cm 0.5cm 0.4cm 0.5cm]{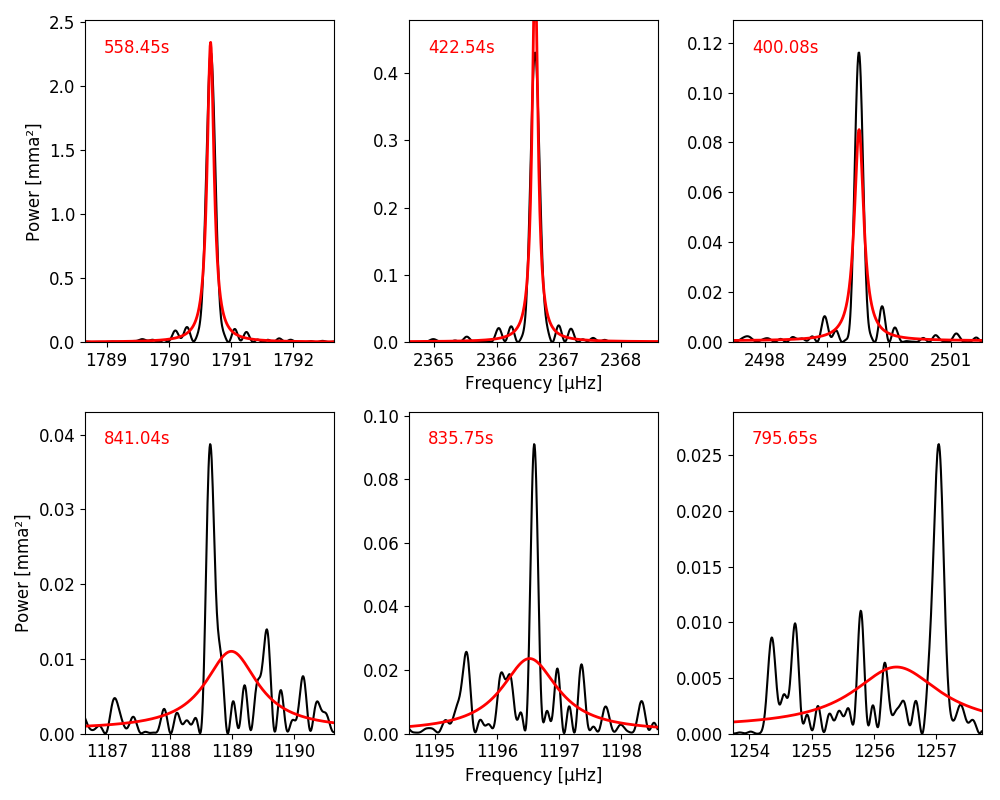}
    \caption{Detailed portions of the {\it K2} data power spectrum of PG~1159-035 illustrating some Lorentzian fits as red lines. The frequency is in $\mu$Hz and the power is in units of mma$^2$ ($1\ \mathrm{mma}=1\ \mathrm{ppt}$). All plots have the same frequency range of $4\,\mu$Hz, but different power ranges. In the top panel, we show some modes whose HWHM are very close to the frequency resolution and, in the bottom panel, we show those modes with the widest HWHM.}
    \label{fig:examples}
\end{figure*}

\par We used the Lorentzian fits to determine the independent frequencies: we assume that all peaks covered by the Lorentzian represent the same mode. And, for modes whose Lorentzian fit covers more than one peak, we defined its HWHM as a width range. For these modes, the uncertainties are unreliable, once they are not coherent over the data set. The frequency and amplitude of the non-coherent modes are, respectively, the central frequency and height of the fitted Lorentzian. 

\par
\begin{figure}
    \centering
    \includegraphics[width=0.52\textwidth,trim=0 0 0 0.5cm]{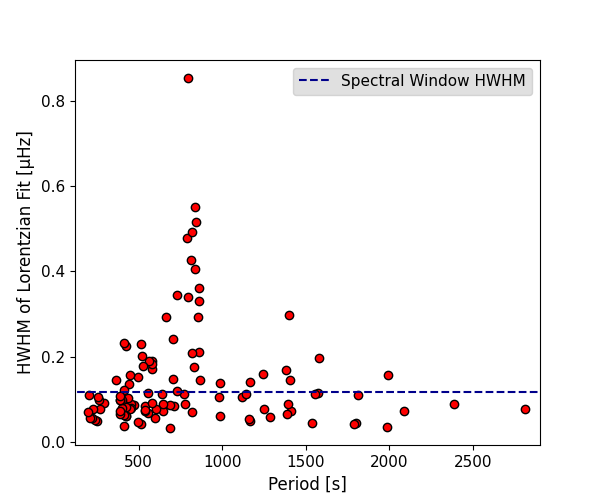}
    \caption{Half-width at half-maximum (HWHM) of Lorentzian fit to the significant peaks in the Fourier Transform of PG~1159-035 in $\mu\mathrm{Hz}$ vs. their periods in seconds. The horizontal blue dashed indicates the spectral window HWHM of the {\it K2} data.}
    \label{fig:HWHM}
\end{figure}

 Some Lorentzian fits are very close to the shape of a single peak, but others cover a few peaks, as illustrated by Figure \ref{fig:examples}.
To determine if the distribution of Lorentzian widths is random or presents some pattern, we plot the HWHM of Lorentzian fits against period. Figure~\ref{fig:HWHM} shows that the largest values of HWHM are in the period range of $\approx 400-1000$~s.  The highest points correspond to three $\ell=1$ modes: $k=33$, 34 and 35 . This is comparable with the results on DAVs of \citet{2017ApJS..232...23H}, who found that the largest values of HWHM are in the period range $800-1400$~s.  We note the largest values of HWHM in their sample of DAVs are several times larger than those seen in the PG~1159-035 data.
\citet{Montgomery20} showed that the dichotomy in HWHM values for the DAVs could be explained by changes in the surface convection zone during pulsation.  These changes alter the reflection condition for modes, making these modes less coherent. 

Models of PG~1159 stars generally show neither surface nor sub-surface convection zones 
\citep[e.g.][]{2006A&A...454..845M,2006A&A...454..863C}\footnote{Only PG~1159 models that have not reached the maximum $T_{\rm eff}$ and evolve towards the blue at constant luminosity, have a thin sub-surface convective zone. However, it disappears completely before reaching the region of interest for PG~1159-035.}, so the mechanism of \citet{Montgomery20} is not expected to lead to a lack of mode coherence for this case. It is possible that other nonlinear effects come in to play near the outer turning point of some modes, which leads to a lack of coherent reflection. 
As an example, the large amplitudes of some modes could lead to a Kelvin-Helmholtz instability (shear instability) in the outer layers of the star, leading to energy loss and inconsistent reflection of the modes. Nonlinear mode coupling on similar timescales has been observed in two pulsating DBVs, by \citet{Kepler03} and \citet{2016A&A...585A..22Z}, consistent with the amplitude equations of \citet{Goupil94} and \citet{1995A&A...296..405B}.

\section{The period spacing and mode identification}
\label{section6}
\subsection{{\it K2} data}
\label{6.1}
\par  According to pulsation theory, the period spacing $\Delta \Pi_{l}$ between two g-modes with the same $\ell$, $m=0$, and consecutive $k$ is constant for a homogeneous model in the asymptotic limit ($k \gg \ell$). We can write the following general equation:
\begin{equation}
    \Pi_{\ell,k} = \Pi_{\ell,0} + k \Delta \Pi_{\ell} 
\end{equation}
where $\Pi_{\ell,k}$ is the period of a $(k,\ell,m=0)$ mode, and $\Pi_{\ell,0}$ is the period of the $k=0$ mode \citep{Tassoul80,Winget91}.

\par
\begin{figure}
    \centering
    \includegraphics[width=0.52\textwidth,trim=0.5cm  0 0 0]{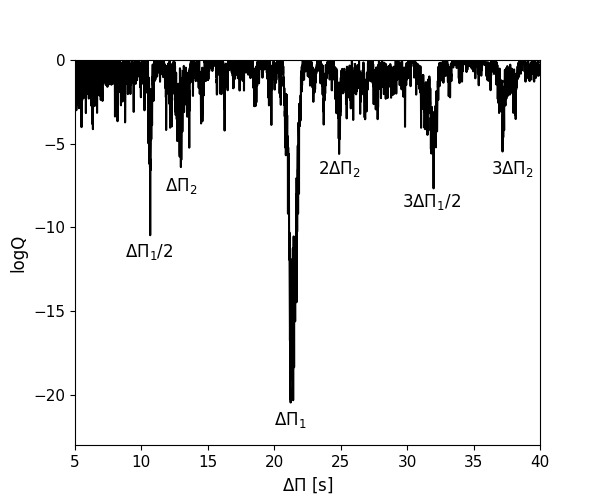}
    \caption{K-S test applied to our list of independent frequencies. There are six significant peaks: two of them correspond to $\Delta \Pi_\mathrm{1}$ and $\Delta \Pi_\mathrm{2}$ values, and the others are multiples of these. }
    \label{K-S}
\end{figure}

 To identify sequences of consecutive {\it k} modes for different $\ell$ values, we need initial guesses for the constants of the equation above. We used the Kolmogorov-Smirnov test (type: KP - Kuiper statistic) to get an initial value for the period spacing $\Delta \Pi_1$ and $\Delta \Pi_2$. Figure~\ref{K-S} shows the K-S test applied to our list of independent frequencies. We identified six significant peaks in the test: $\Delta \Pi_\mathrm{1} = 21.23$~s, $\Delta \Pi_{\mathrm{2}} = 12.97$~s and two multiples of each one. The $\Delta \Pi_\mathrm{1}$ peak is significantly stronger than its multiples. This is not true to $\Delta \Pi_\mathrm{2}$, which has a significance $\log Q$ value similar to its multiples. This may result from the smaller number of $\ell=2$ frequencies observed compared to the sample of $\ell=1$ frequencies.

\par We would also like to place constraints on the values of $\bf \Pi_{\ell=1,0}$ and $\bf \Pi_{\ell=2,0}$.  For slow rotation in the absence of a magnetic field, the components of a split mode are equally spaced in frequency by a constant  $\delta \nu_{\ell}$. To identify multiplets in the set of frequencies observed, it is important to estimate $\delta \nu_{\ell}$. As initial guesses, we took the values obtained  by \citet{Winget91}, $\delta \nu_{\ell=1} = 4.22$ and $\delta \nu_{\ell=2}=6.92$. Further, using the results of \citet{Winget91}  to assume that the triplet at $517$~s is the $k=20 \pm 2$\footnote{The radial overtone number $k$ is constrained from theory to within ~$\pm 2$ \citet{Winget91}},\ $\ell=1$ mode, we find from equation $(1)$: $\Pi_{0,\ell=1} = 92.6$~s.  Using the results of \citet{Costa08} to assume that the quintuplet of $387$ s is the $k=27 \pm 2,\ \ell=2$ mode, we obtain $\Pi_{0,\ell=2} = 37.98$~s. 
\par Taking these values to classify the {\it K2} frequencies in the range of $300-3000$ $\mu \mathrm{Hz}$, we identified $32$ $\ell=1$ independent modes (Table \ref{tab:l=1}) and $12$ $\ell=2$ independent modes (Table \ref{tab:l=2}). The fourteen $\ell=1$ modes with the largest values of $k$ are out of the frequency range analyzed by \citet{Costa08}, and the $k=61,\ell=2$ mode was not detected in their data.

\begin{table*}[t]
    \centering
    \begin{tabular}{|ccclc|ccclc|}\hline
        $k \pm 2$ & m & Period & Frequency & Amplitude & $k \pm 2$  & m & Period & Frequency & Amplitude\\
         & & [s] &  [$\mu$Hz] &  [mma] &   &  &  [s] &  [$\mu$Hz] &  [mma]  \\ \hline \hline
        
14  &  +1  & $ 386.8818 \pm 0.0013$ & $2584.769 \pm 0.008$ & $0.26 \pm 0.02$ &     35*   &  +1  & $ 835.75 \pm 0.38$ & $1196.53 \pm 0.55 $ & $0.15$\\
14  &  0  & $ 387.5045 \pm 0.0013$ & $2580.615 \pm 0.009$ & $0.25 \pm 0.02$ &      35*   &  0   &  $ 838.50 \pm 0.29$ & $1192.61\ \pm 0.41 $ & $0.13$\\
14  &  -1   &          &                      &          &   35*   &  -1  & $ 841.05 \pm 0.36$ & $1188.99 \pm 0.52$ & $0.10$\\ \hline
17  &  +1  & $ 451.6086 \pm 0.0002$ & $2214.307 \pm 0.001$ & $2.70 \pm 0.02$ &     42   &  +1  & $ 981.7223 \pm 0.0100$ & $1018.618 \pm 0.010$ & $0.21 \pm 0.02$\\
17  &  0   & $ 452.4465 \pm 0.0002$ & $2210.206 \pm 0.001$ & $1.87 \pm 0.02$ &      42   &  0   & $ 985.6528 \pm 0.0093$ & $1014.556 \pm 0.010$ & $0.23 \pm 0.02$\\
17  &  -1  & $ 453.2911 \pm 0.0001$ & $2206.088 \pm 0.001$ & $3.14 \pm 0.02$ &      42   &  -1  & $ 988.4763 \pm 0.0011$ & $1011.658 \pm 0.001$ & $1.88 \pm 0.02$\\ \hline
18  &  +1  &          &                      &        &     48   &  +1  &           &                     &        \\
18  &  0   & $ 473.0622 \pm 0.0009$ & $2113.887 \pm 0.004$ & $0.52 \pm 0.02$ &      48   &  0   & $1116.0179 \pm 0.0096$ & $ 896.043 \pm 0.008$ & $0.29 \pm 0.02$\\
18  &  -1  &          &                      &        &    48   &  -1  &           &                    &        \\ \hline
20*  &  +1  &  $ 516.10 \pm 0.06$  & $1937.59\ \pm 0.23 $ & $1.28$ &     50   &  +1  & $1159.3618 \pm 0.0085$ & $ 862.543 \pm 0.006$ & $0.35 \pm 0.02$\\
20  &   0  & $ 517.2150 \pm 0.0001$ & $1933.432 \pm 0.001$ & $5.34 \pm 0.02$ &     50   &  0   & $1164.8798 \pm 0.0107$ & $ 858.458 \pm 0.008$ & $0.28 \pm 0.02$\\
20*  &  -1  & $ 518.34 \pm 0.05$  & $1929.24\ \pm 0.20 $ & $1.46$ &     50   &  -1  & $1168.7091 \pm 0.0088$ & $ 855.645 \pm 0.006$ & $0.34 \pm 0.02$\\ \hline
21  &  +1  & $ 536.9727 \pm 0.0001$ & $1862.292 \pm 0.001$ & $5.92 \pm 0.02$ &   54   &  +1  & $1242.0046 \pm 0.0150$ & $ 805.150 \pm 0.010$ & $0.23 \pm 0.02$\\
21  &  0   & $ 538.1587 \pm 0.0002$ & $1858.188 \pm 0.001$ & $4.08 \pm 0.02$ &   54   &  0   & $1246.2601 \pm 0.0216$ & $ 802.401 \pm 0.014$ & $0.16 \pm 0.02$\\
21  &  -1  & $ 539.3473 \pm 0.0001$ & $1854.093 \pm 0.001$ & $4.74 \pm 0.02$ &   54   &  -1  & $1252.6274 \pm 0.0178$ & $ 798.322 \pm 0.011$ & $0.19 \pm 0.02$\\ \hline
22  &  +1  & $ 557.1217 \pm 0.0006$ & $1794.940 \pm 0.002$ & $1.14 \pm 0.02$ &    56   &  +1  &           &                    &        \\
22  &  0   & $ 558.4483 \pm 0.0004$ & $1790.676 \pm 0.001$ & $1.53 \pm 0.02$ &    56   &  0   & $1284.5235 \pm 0.0254$ & $ 778.499 \pm 0.015$ & $0.14 \pm 0.02$\\
22*  &  -1  &  $ 559.78 \pm 0.06$  & $1786.42 \pm 0.19 $ & $0.27$ &      56   &  -1  &           &                   &        \\ \hline
23*  &  +1  &  $ 579.07 \pm 0.06$  & $1726.90 \pm 0.18 $ & $0.11$ &      61   &  +1  & $1379.2645 \pm 0.0309$ & $ 725.024 \pm 0.016$ & $0.14 \pm 0.02$\\
23*  &  0   &  $ 580.38 \pm 0.06$  & $1723.00 \pm 0.19 $ & $0.18$ &      61   &  0   & $1387.0651 \pm 0.0210$ & $ 720.947 \pm 0.011$ & $0.20 \pm 0.02$\\
23*  &  -1  & $581.78 \pm 0.06$ &	$1718.85 \pm 0.17$ &	$0.20$ &  61   &  -1  & $1392.6642 \pm 0.0299$ & $ 718.048 \pm 0.015$ & $0.14 \pm 0.02$\\ \hline
24  &  +1  & $ 600.6547 \pm 0.0023$ & $1664.850 \pm 0.006$ & $0.34 \pm 0.02$ &  62*   &  +1  & $1398.74 \pm 0.58$ & $ 714.93 \pm 0.30$ & $0.16$\\
24  &  0   &          &                      &        &   62   &  0   & $1406.5058 \pm 0.0122$ & $ 710.982 \pm 0.006$ & $0.36 \pm 0.02$\\
24  &  -1  & $ 604.0409 \pm 0.0041$ & $1655.517 \pm 0.011$ & $0.20 \pm 0.02$ &     62   &  -1  & $1412.1701 \pm 0.0000$ & $ 708.130 $ & $0.72 \pm 0.02$\\ \hline
26  &  +1  & $ 641.5557 \pm 0.0006$ & $1558.711 \pm 0.001$ & $1.49 \pm 0.02$ &     68   &  +1  &           &                   &        \\
26  &  0   & $ 643.2361 \pm 0.0004$ & $1554.639 \pm 0.001$ & $2.54 \pm 0.02$ &     68   &  0   & $1539.0279 \pm 0.0342$ & $ 649.761 \pm 0.014$ & $0.15 \pm 0.02$\\
26  &  -1  & $ 644.9279 \pm 0.0002$ & $1550.561 \pm 0.001$ & $3.68 \pm 0.02$ &     68   &  -1  &           &                    &         \\ \hline
27  &  +1  &          &                      &        &     69   &  +1  &          &                   &        \\
27  &  0   & $ 664.1985 \pm 0.0068$ & $1505.574 \pm 0.015$ & $0.14 \pm 0.02$ &   69   &  0   & $1555.7422 \pm 0.0335$ & $ 642.780 \pm 0.014$ & $0.16 \pm 0.02$\\
27  &  -1  &          &                      &        &     69   &  -1  &           &                    &        \\ \hline
28  &  +1  & $ 685.7765 \pm 0.0067$ & $1458.201 \pm 0.014$ & $0.15 \pm 0.02$ &    70   &  +1  &           &                    &       \\ 
28  &  0   &          &                      &        &      70*   &  0   & $1572.9948 \pm 0.0000$ & $ 635.730 $ & $0.37 \pm 0.02$\\
28  &  -1  & $ 689.6951 \pm 0.0056$ & $1449.916 \pm 0.012$ & $0.19 \pm 0.02$ &     70   &  -1  &  $1589.91 \pm 0.49$  & $ 632.95 \pm 0.20 $ & $0.29$\\\hline
29*  &  +1  &  $ 706.08 \pm 0.12$  & $1416.27 \pm 0.24 $ & $0.27$ &     80   &  +1  & $1788.9168 \pm 0.0392$ & $ 558.998 \pm 0.012$ & $0.18 \pm 0.02$\\
29  &  0   & $ 708.1235 \pm 0.0021$ & $1412.183 \pm 0.004$ & $0.51 \pm 0.02$ &     80   &  0   & $1802.1048 \pm 0.0437$ & $ 554.907 \pm 0.013$ & $0.16 \pm 0.02$\\
29  &  -1  & $ 710.3398 \pm 0.0018$ & $1407.777 \pm 0.004$ & $0.61 \pm 0.02$ &     80   &  -1  & $1811.2402 \pm 0.0400$ & $ 552.108 \pm 0.012$ & $0.18 \pm 0.02$\\ \hline
30  &  +1  &          &                      &        &     89   &  +1  &           &                   &  \\
30*  &  0   & $ 729.60 \pm 0.18$ & $1370.61 \pm 0.35$ & $0.11$ &    89   &  0   & $1982.7768 \pm 0.0504$ & $ 504.343 \pm 0.013$ & $0.17 \pm 0.02$\\
30  &  -1  & $ 731.6054 \pm 0.0037$ & $1366.857 \pm 0.007$ & $0.32 \pm 0.02$ &     89   &  -1  &  $1993.56 \pm 0.07$ & $501.615 \pm 0.017$ & $0.13 \pm 0.02$  \\ \hline
32  &  +1  &          &                      &        &    90   &  +1  &           &                   & \\
32  &  0   & $ 773.7070 \pm 0.0066$ & $1292.479 \pm 0.011$ & $0.20 \pm 0.02$ &    90   &  0   & $2010.0490 \pm 0.0569$ & $ 497.500 \pm 0.014$ & $0.16 \pm 0.02$\\
32  &  -1  & $ 776.0878 \pm 0.0078$ & $1288.514 \pm 0.013$ & $0.17 \pm 0.02$ &    90   &  -1  & $2021.2441 \pm 0.0613$ & $ 494.745 \pm 0.015$ & $0.15 \pm 0.02$\\ \hline
33*  &  +1  &  $ 791.56 \pm 0.30$  & $1263.33\ \pm 0.48 $ & $0.16$ &    94   &  +1  &           &                   &  \\
33*  &  0   &  $ 793.89 \pm 0.21$  & $1259.62 \pm 0.34 $ & $0.20$ &     94   &  0   & $2084.6910 \pm 0.0501$ & $ 479.687 \pm 0.012$ & $0.19 \pm 0.02$\\
33*  &  -1  & $ 795.5298 \pm 0.0092$ & $1257.024 \pm 0.015$ & $0.15 \pm 0.02$ &    94   &  -1  &           &                    & \\ \hline
34*  &  +1  & $ 813.77 \pm 0.28$ & $1228.85 \pm 0.43$ & $0.08$ &    128   &  +1  &           &                    &  \\
34*  &  0   &  $ 816.68 \pm 0.14$ & $1224.46 \pm 0.21 $ & $0.13$ &   128   &  0   & $2807.9405 \pm 0.0463$ & $ 356.133 \pm 0.006$ & $0.37 \pm 0.02$\\
34*  &  -1  & $ 819.02\pm 0.33$ & $1220.98 \pm 0.49$ & $0.15$ &   128   &  -1  &           &                   & \\ \hline

    \end{tabular}
    \caption{Identified $\ell=1$ pulsation modes in {\it K2} data. The frequencies with an `*' after the k number are not coherent during observations, and their parameters are the values referring to Lorentzian fits.}
    \label{tab:l=1}
\end{table*}

\begin{table*}
    \centering
    \begin{tabular}{|ccclc|ccclc|}\hline
        $k \pm 2$ & m & Period & Frequency & Amplitude & $k \pm 2$  & m & Period & Frequency & Amplitude\\
         & & [s] &  [$\mu$Hz] &  [mma] &   &  &  [s] &  [$\mu$Hz] &  [mma]  \\ \hline \hline

25  &  +2  &          &                      &        &    32  &  +2  &          &                      &       \\
25  &  +1  &          &                      &        &    32  &  +1  &          &                      &      \\
25  &   0  & $ 362.5385 \pm 0.0014$ & $2758.328 \pm 0.011$ & $0.21 \pm 0.02$ &   32  &   0  & $ 449.3623 \pm 0.0032$ & $2225.376 \pm 0.016$ & $0.14 \pm 0.02$\\
25  &  -1  &          &                      &        &   32  &  -1  &          &                      &       \\
25  &  -2  &          &                      &        &    32  &  -2  &          &                      &       \\ \hline
27  &  +2  &          &                      &        &  36  &  +2  &          &                      &       \\
27  &  +1  & $ 387.1166 \pm 0.0005$ & $2583.201 \pm 0.003$ & $0.65 \pm 0.02$ &    36  &  +1  & $ 497.0416 \pm 0.0028$ & $2011.904 \pm 0.011$ & $0.19 \pm 0.02$\\
27  &   0  &          &                      &         &  36  &   0  & $ 498.7392 \pm 0.0011$ & $2005.056 \pm 0.004$ & $0.50 \pm 0.02$\\
27  &  -1  & $ 389.2308 \pm 0.0002$ & $2569.170 \pm 0.002$ & $1.45 \pm 0.02$ &   36  &  -1  &          &                      &      \\
27  &  -2  & $ 390.2865 \pm 0.0002$ & $2562.220 \pm 0.001$ & $2.12 \pm 0.02$ &   36  &  -2  &          &                      &      \\ \hline
28  &  +2  &          &                      &         &  38  &  +2  &          &                      &       \\
28  &  +1  & $ 398.9886 \pm 0.0013$ & $2506.337 \pm 0.008$ & $0.28 \pm 0.02$ &   38  &  +1  &          &                      &       \\
28  &   0  & $ 400.0776 \pm 0.0010$ & $2499.515 \pm 0.006$ & $0.34 \pm 0.02$ &   38*  &   0  & $ 528.21 \pm 0.05$ & $1893.20\ \pm  0.18$ & $0.11$\\
28  &  -1  &          &                      &        &  38  &  -1  &          &                      &       \\
28  &  -2  &          &                &   &  38  &  -2  &          &                      &      \\ \hline
29  &  +2  & $ 410.8695 \pm 0.0025$ & $2433.863 \pm 0.015$ & $0.15 \pm 0.02$ &    61  &  +2  &    &   &  \\
29*  &  +1  & $412.04 \pm 0.04$ &	$2426.97 \pm 0.23$ &	$0.24$ &   61  &  +1  &          &                      &      \\
29  &   0  &          &                  &   &  61  &   0  &          &                      &       \\
29  &  -1  & $ 414.4032 \pm 0.0004$ & $2413.109 \pm 0.002$ & $1.05 \pm 0.02$ &   61*  &  -1  & $ 833.58 \pm 0.12$ & $1199.65 \pm 0.18$ & $0.17$\\
29  &  -2  & $ 415.5943 \pm 0.0008$ & $2406.193 \pm 0.005$ & $0.48 \pm 0.02$ &   61  &  -2  &          &                      &       \\ \hline
30  &  +2  & $ 422.5439 \pm 0.0006$ & $2366.618 \pm 0.003$ & $0.65 \pm 0.02$ &  63  &  +2  &          &                      &       \\
30  &  +1  & $ 423.7669 \pm 0.0004$ & $2359.788 \pm 0.002$ & $0.94 \pm 0.02$ &   63  &  +1  &          &                      &       \\
30  &   0  &             &        &   &  63*  &   0  & $856.55 \pm 0.21$ &	$1167.47 \pm 0.29$ &	$0.28$ \\
30*  &  -1  & $426.26 \pm 0.04$ & $2346.01 \pm 0.22$ &	$0.19$ &   63*  &  -1  & $ 861.00 \pm 0.25$ & $1161.44 \pm 0.33$ & $0.11$\\
30  &  -2  & $ 427.4673 \pm 0.0017$ & $2339.360 \pm 0.009$ & $0.23 \pm 0.02$ &  63  &  -2  & $ 866.2231 \pm 0.0081$ & $1154.437 \pm 0.011$ & $0.20 \pm 0.02$\\ \hline
31  &  +2  &          &                      &       &  64*  &  +2  & $859.36 \pm 0.27$ &	$1163.66 \pm 0.36$ &	$0.14$ \\
31  &  +1  &          &                      &      &  64*  &  +1  & $864.21 \pm 0.16$ &	$1157.13 \pm 0.21$ &	$0.20$ \\
31  &   0  &     &        &   &  64  &   0  &          &                      &       \\
31  &  -1  & $ 439.3573 \pm 0.0007$ & $2276.052 \pm 0.004$ & $0.57 \pm 0.02$ & 64  &  -1  &          &                      &       \\
31  &  -2  & $ 440.7055 \pm 0.0021$ & $2269.089 \pm 0.011$ & $0.20 \pm 0.02$ & 64  &  -2  &          &                      &       \\\hline

    \end{tabular}
    
    \caption{Identified $\ell=2$ pulsation modes in {\it K2} data. The frequencies with an `*' after the k number are not coherent during observations, and their parameters are the values referring to Lorentzian fits.}
    \label{tab:l=2}
\end{table*}

\begin{figure}
    \centering
    \includegraphics[width = 0.52\textwidth,trim=0.5cm 0 0 0.5cm]{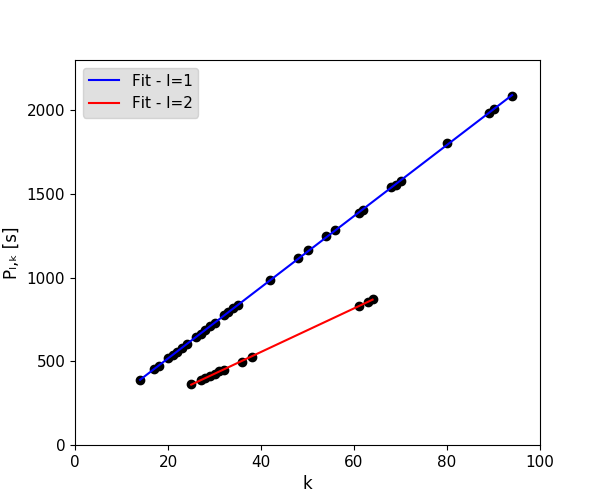}
    \caption{Observed periods sequences for $\ell=1$ (blue line) and $\ell=2$ (red line) modes from {\it K2} data.}
    \label{P_k}
\end{figure}

\par After identifying the modes, we were able to use their observational values to improve our constants. We plotted the periods vs. their assigned $k$ value to the $\ell=1$ and $\ell=2$ modes, as shown in 
Figure~\ref{P_k}. So, fitting a line according to Equation~$(1)$, we obtain:
\begin{equation}
\Delta \Pi_{\textit{1}} = 21.28 \pm 0.02~\mathrm{s} \quad \mathrm{and} \quad 
\Delta \Pi_{\textit{2}} = 13.02 \pm 0.04~\mathrm{s}
\end{equation}
These values are very similar to the ones found by \citet{Costa08}: $\Delta \Pi_{\textit{1}} = 21.43 \pm 0.03~\mathrm{s}$ and $\Delta \Pi_{\textit{2}} = 12.38 \pm 0.01~\mathrm{s}$. The ratio between the period spacings we obtained is:
\begin{equation}
    \frac{\Delta \Pi_{\textit{1}}}{\Delta \Pi_{\textit{2}}} = 1.634 \pm 0.005 
\end{equation}
that is, 94\% of $\sqrt{3}$, the value expected by asymptotic pulsation theory. 

\subsection{{\it TESS} data}
\label{tess-data}
\par In Fig.~\ref{FT+SW} we observe that {\it TESS} data is much noisier than {\it K2} data. The higher noise levels make it impossible to detect low amplitude frequencies.  We are unable to complete a deep analysis. Since we do not gain any new insights from the {\it TESS} data compared to the {\it K2} data, we only consider the {\it K2} data for the remainder of this manuscript.

\begin{table}[ht!]
    \centering
 
    \begin{tabular}{|ccccc|} \hline
         $k \pm 2$ & m & Period & Frequency & Amplitude\\
         & & [s] &  [$\mu$Hz] &  [mma] \\ \hline \hline
        17	&	 1	&                           &                         &                \\
        17	&	 0	&	  $452.462 \pm 0.006$	&	  $2210.13 \pm 0.03$  &	 $3.7 \pm 0.4$ \\
        17	&	-1	&	  $453.307 \pm 0.009$	&	  $2206.01 \pm 0.04$  &	 $2.4 \pm 0.4$ \\ \hline
        20	&	 1	&	  $516.107 \pm 0.006$	&	  $1937.58 \pm 0.02$  &  $4.5 \pm 0.4$ \\
        20	&	 0	&	  $517.211 \pm 0.005$	&	  $1933.45 \pm 0.02$  &	 $5.7 \pm 0.4$ \\
        20	&	-1	&	  $518.370 \pm 0.005$	&	  $1929.12 \pm 0.02$  &  $5.2 \pm 0.4$ \\ \hline
        21	&	 1	&		&	   &   \\
        21	&	 0	&	  $538.155 \pm 0.006$   &	  $1858.20 \pm 0.02$  &	 $5.5 \pm 0.4$ \\
        21	&	-1	&	  $539.333 \pm 0.005$	&	  $1854.14 \pm 0.02$  &	 $5.5 \pm 0.4$ \\ \hline
    \end{tabular}

    \caption{Identified pulsation frequencies in {\it TESS} data.}
    \label{tab:tess}
\end{table}


\section{Mode structure and Asymmetries}
\label{section7}
\subsection{$\ell = 1$ modes}
\par
\begin{figure}
    \centering
    \includegraphics[width = 0.4\textwidth, trim = 0cm 0.5cm 0cm 0.5cm]{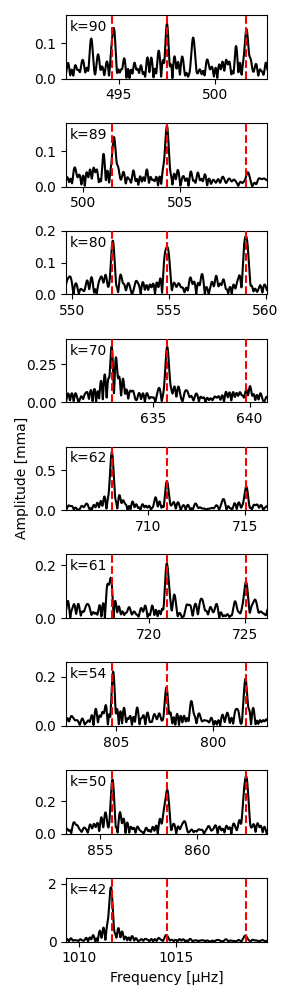}
    \caption{The panels show detailed portions of the {\it K2} FT centered in the $m=0$ frequency of the asymmetric modes. The red dashed lines indicates the $\langle \delta \nu_\mathrm{rot,1} \rangle$ frequency spacing averages.}
    \label{asymmodes}
\end{figure}

 The 32 $\ell=1$ modes identified in the {\it K2} light curve are distributed in 8 singlets, 8 doublets, and 16 triplets. For the 24 modes with multiplet structure, we find 15 with approximately symmetric splitting of  $\langle \delta \nu_{\rm rot,1} \rangle = (4.0 \pm	0.4)~\mu$Hz.  The
remaining nine $\ell=1$ multiplets have a very similar asymmetric frequency structure, as shown in Figures~\ref{asymmodes} and \ref{rotation}. 
The asymmetric modes present a larger spacing average $\langle \delta \nu_{\rm rot,1}^{+} \rangle = (4.06 \pm 0.05)~\mu$Hz and a smaller spacing average  $\langle \delta \nu_{\rm rot,1}^{-} \rangle = (2.81 \pm 0.06)~\mu$Hz.  For the major asymmetric modes, the larger spacing is that between the $m=0$ and $m=1$ frequencies.  The $k=54$ mode
is an exception, with the larger spacing between the $m=-1$ and $m=0$ frequencies.  Please note that the x-axis of the $k=54$ panel is inverted in Figure~\ref{asymmodes}. 

\par
\begin{figure}
    \centering
    \includegraphics[width = 0.5\textwidth, trim= 0.5cm 0 0 1.4cm]{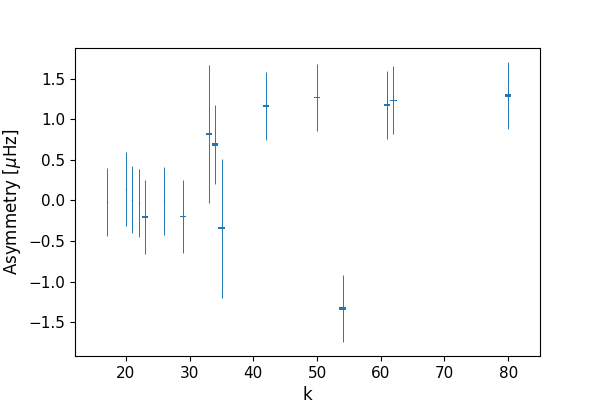}
    \caption{Asymmetry of the $\ell$ = 1 modes  as a function of the radial node $k$. The height of the square markers     correspond to the internal uncertainties calculated by MC Monte Carlo simulation. The error bars indicate the uncertainties obtained by the Lorentzian fit.}
    \label{asymmetry}
\end{figure}

  Figure \ref{asymmetry} shows the asymmetry observed in $\ell = 1$ modes in PG~1159$-$035 as a function of the radial node $k$. While the pulsations are global, each pulsation samples the star in slightly different ways. The lower {\it k} modes have outer turning points that are far below the stellar surface, so these modes preferentially sample the deeper interior.  High {\it k} modes sample more of  the outer layers.  
Figure~\ref{asymmetry} shows that the larger asymmetries are found in modes with larger values of {\it k}. This argues that the cause of the asymmetries, whether it be magnetic fields, differential rotation or some other symmetry-breaking agent, predominately influences the outer layers corresponding to $40 \leq k \leq 80$  rather than in the inner zone corresponding to $k \leq 30$.

\par In the case of weak magnetic fields and slow ($\omega_\mathrm{puls}>\Omega_\mathrm{rot}$) uniform rotation, the observed frequency spacing depends on both the rotation period and the magnetic field of the star. Assuming that  the pulsation axis, the rotation axis and the magnetic field are aligned, the frequency spacing can be approximated in first order by \citep[e.g.][]{1977ApJ...217..151H, 1989ApJ...336..403J,Gough90,Dintrans00} :
\begin{equation}
    \delta \nu\ \approx\ m(1-C)\Omega_\mathrm{rot}\ +\ m^2 \bar{\gamma}B^2 
    \label{deltanu}
\end{equation}
where the Ledoux constant $C=C(k,\ell)$ is the uniform rotation coefficient \citep{Ledoux51}, $\Omega_\mathrm{rot}$ is the rotational frequency, $\bar{\gamma}$ is a proportionality constant and $B$ is the magnetic field. Because of its dependence on $m^2$, the magnetic field term in equation \ref{deltanu} causes an asymmetric splitting about $m = 0$ if combined with the rotational effect. Defining: 
$$\mathrm{Asymmetry}(k,\ell) \approx \delta \nu(k,\ell,m=+1) + \delta \nu(k,\ell,m=-1)$$
\begin{equation}
 \mathrm{Asymmetry}(k,\ell) \approx  2 \bar{\gamma}B^2 
\end{equation}
with the asymmetry directly related with the square of the magnetic field. 

From Figure~\ref{rotation}, however, we can see that the exact morphology of the splitting does not follow the form of equation~\ref{deltanu}. While the $m=-1$ components are shifted to the right (toward the $m=0$ modes), the $m=+1$ components should \emph{also} be shifted to the right, away from the $m=0$ modes. This is not observed. Thus, the simple model of uniform rotation plus a uniform magnetic field in which the magnetic and rotation axes are aligned cannot be valid.
But if we assume differential rotation, as suggested by Figure~\ref{rotation}, a lower rotational frequency in the period formation zone of the asymmetric modes plus an off-center magnetic field could explain Figure~\ref{asymmetry}. Second order effects of rotation also can produce an asymmetry in multiplets, even for the slow rotators \citep{1992ApJ...394..670D}.

\par
\begin{figure}
    \centering
    \includegraphics[width=0.52\textwidth, trim= 0.7cm 0 0cm 1.4cm]{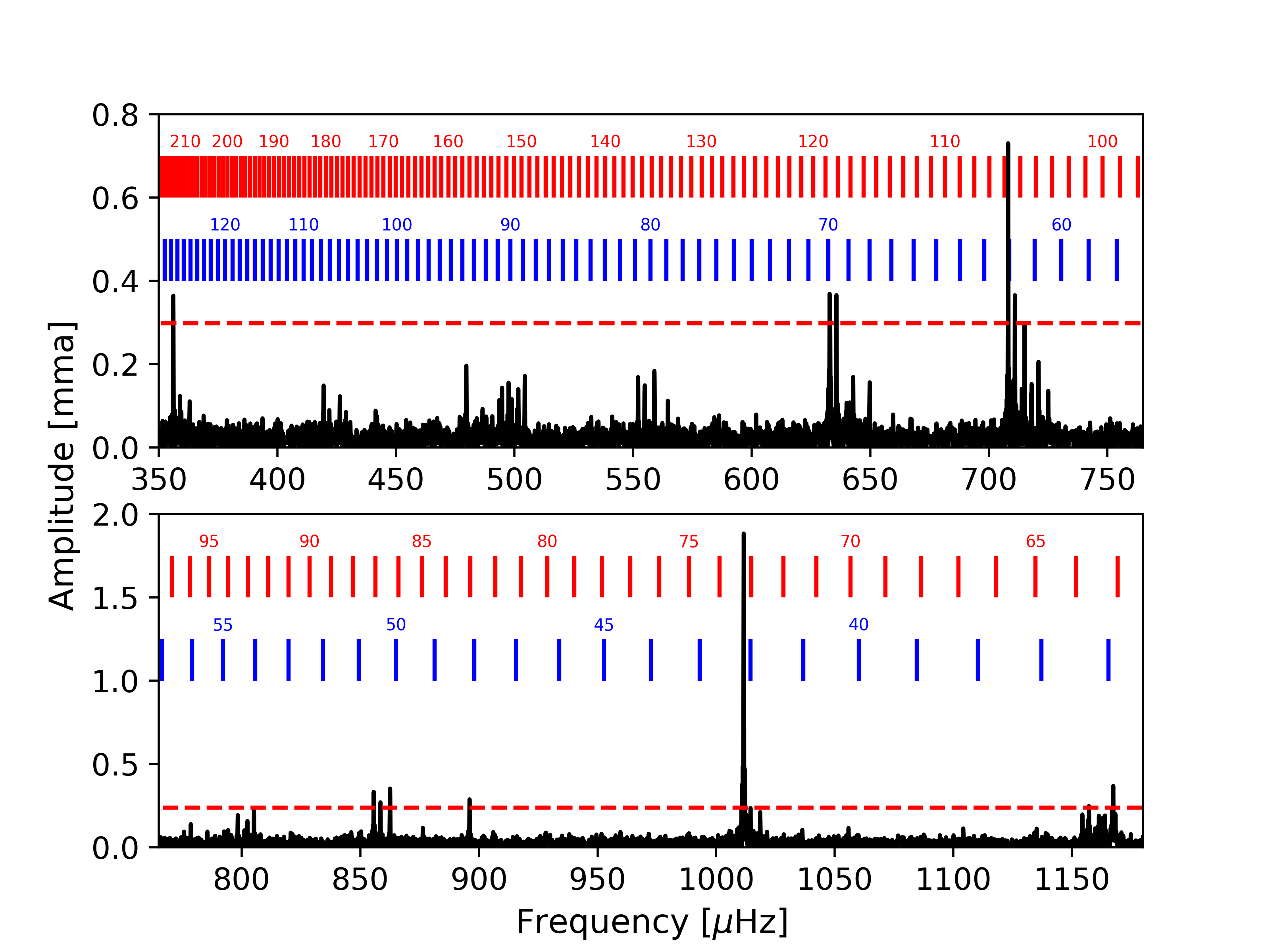}
    \caption{K2 data FT in the range of asymmetric modes. Vertical blue lines indicate the component m=0 of the $\ell=1$ model calculated in Section \ref{6.1}, similarly, the red ones indicate the component m=0 of the $\ell=2$ model. Red horizontal dashed line indicates the fap limit.}
    \label{fig:overlap}
\end{figure}

 However, it is important to point that all this discussion is based on the assumption that the quantum numbers $\it{k}$, $\ell$ and $m$ assigned to the frequencies lower than 1100 $\mu$Hz are correct.  We have suggested a mode identification in Table~\ref{tab:l=1} using the $\Pi_{\ell,k}$ model we have calculated in section \ref{6.1}. In the range of low frequencies, this model present a strong superposition of both $\ell=1$ and $\ell=2$ modes (see Figure~\ref{fig:overlap}), leading to other equally possible mode identification to these frequencies: the possibility that these sets of frequencies are not asymmetric $\ell=1$ modes but superposition of $\ell=1$ and $\ell=2$ modes, or nonlinear difference frequencies. 

\subsection{$\ell = 2$ modes}
\par Our sample for $\ell=2$ contains 12 modes distributed in 3 singlets, and 9 multiplets. We do not find any $\ell=2$ mode where all five $m$ subcomponents are detected.  Based on this small sample, we did not observe any pattern of asymmetry. The frequency spacing average and standard deviation are: 
\begin{equation}
    \langle\delta \nu_\mathrm{rot,2} \rangle = 6.8 \pm 0.2~\mu\mathrm{Hz}
\end{equation}
As here we used only the {\it K2} data, both the values of frequency spacing average we estimated are less accurate than the multiyear values calculated by \citet{Costa08}: $ \langle\delta \nu_\mathrm{rot,1} \rangle = 4.134 \pm 0.002~\mu\mathrm{Hz}$ and $ \langle\delta \nu_\mathrm{rot,2} \rangle = 6.90 \pm 0.01~\mu\mathrm{Hz}$.

\subsection{Regions of period formation}

\begin{figure}
    \centering
    \includegraphics[width=0.45\textwidth, trim= 0.5cm 0 0 0]{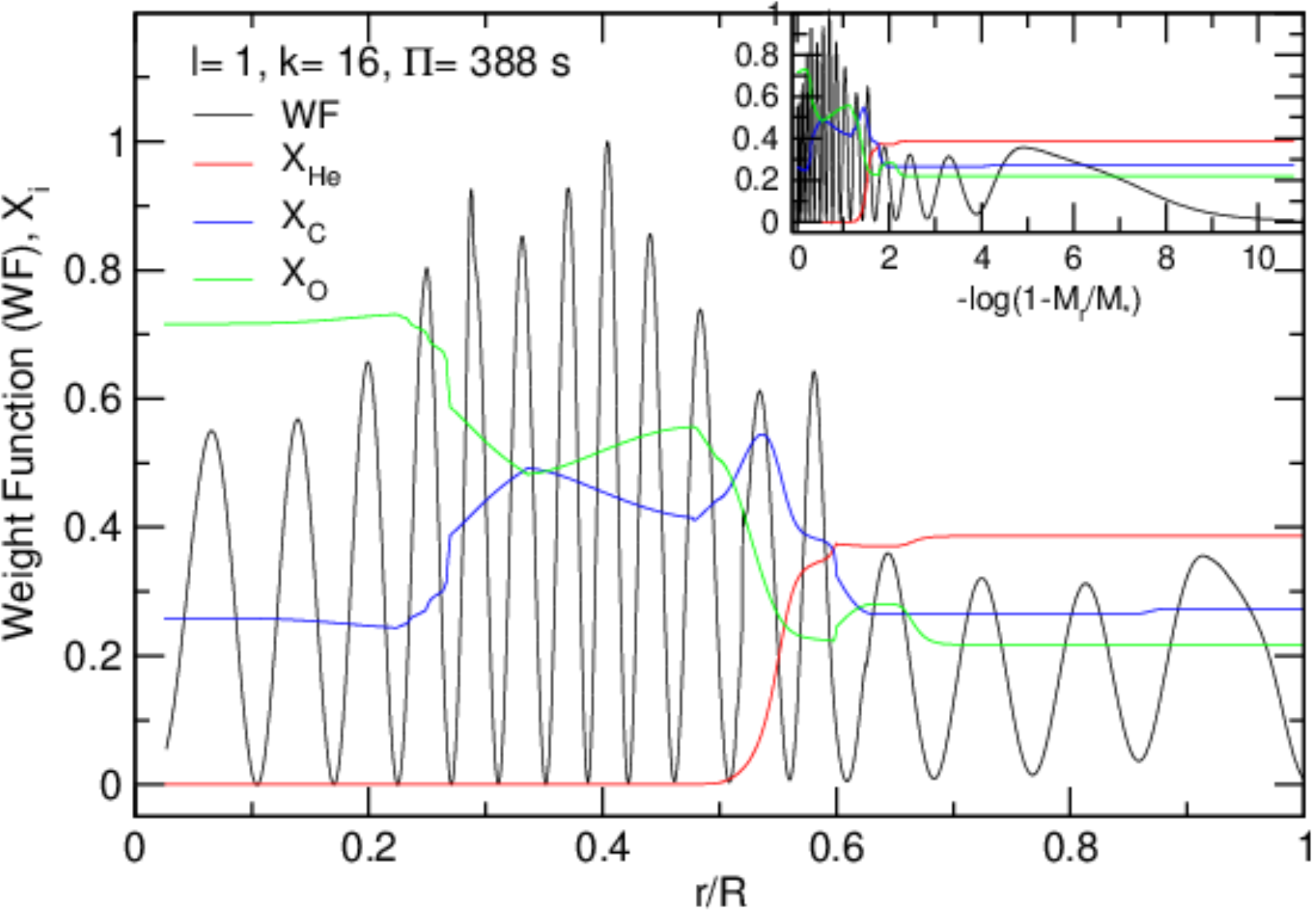}
    \caption{Normalized weight function of the $k=16$ mode (in  black) 
    versus the normalized stellar radius. The chemical composition profiles are also indicated by lines with different colors. Inset: the same quantities as in the main figure, but as a function of the negative logarithm of the outer mass fraction.}
    \label{fig:k16}
\end{figure}

\begin{figure}
    \centering
    \includegraphics[width=0.45\textwidth, trim= 0.5cm 0 0 0]{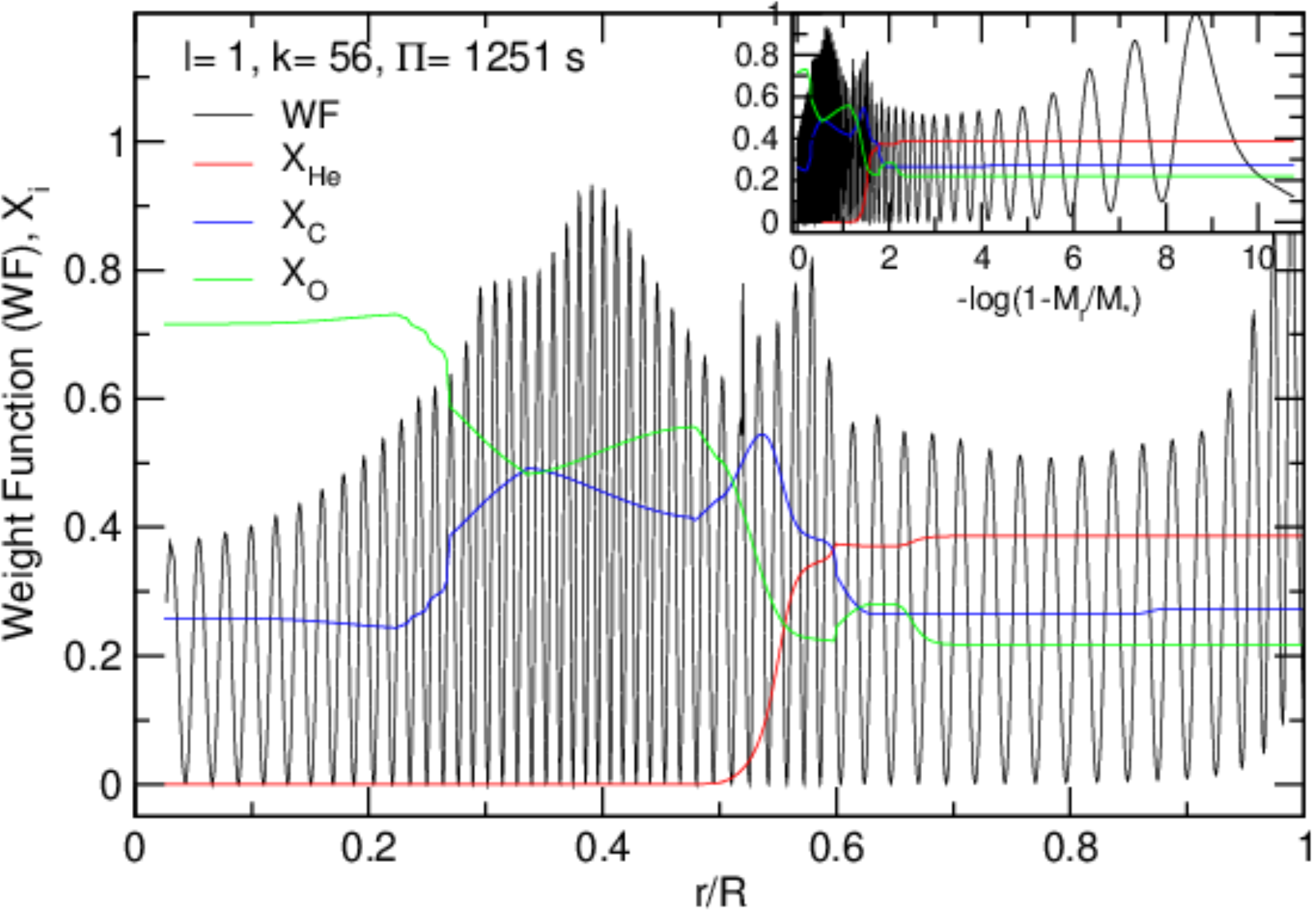}
    \caption{Same as fig. \ref{fig:k16}, but for the $k= 56$ mode.}
    \label{fig:k56}
\end{figure}

Figures~\ref{fig:k16} and \ref{fig:k56} show the normalized ``weight functions'' versus the normalized radius for the $k=16$ mode and for the $k=56$ mode, computed as in \cite{1985ApJ...295..547K}, for a representative PG~1159 model. The inset of each Figure depicts the same functions but in terms of the outer mass fraction coordinate. For reference, the internal chemical abundances are 
also shown. The weight functions 
constitute a very useful diagnostic recipe to know which are the most 
relevant regions of the star for the formation of periods. Due to 
the large $k$ values, there are no large differences in the regions 
of period formation for the modes with $k=16$ and $k=56$.

\section{The rotational period}
\label{section8}
\par At first glance, the rotation period of PG1159-035 appears well established.  Analysis of the {\it K2} modes find an average splitting of $\delta\nu\ = 4.0\pm0.4\ \mu$Hz for $\ell = 1$, and $\delta\nu\ = 6.8\pm0.2\ \mu$Hz for $\ell= 2$.  Using the equations for uniform rotation in the asymptotic limit, we find a rotation frequency of 
$8.16\ \mu$Hz and a period of $P_{\rm rot}=1.4\pm0.1$ days.  This is in agreement with the $1.38\pm0.013$ day rotation period reported in \citet{Winget91} and $1.3935\pm0.0008$ day period reported in \citet{Costa08}.  
\par The Fourier Tranform of {\it K2} light curve reveals a significant ($1.46$~mma) peak at $8.906\pm0.003\ \mu\mathrm{Hz}$ as well as its harmonic at $17.811\pm0.006\,\mu$Hz. The modulation is apparent in visual inspection of the {\it K2} light curve. There is no evidence that PG~1159-035 is a member of a binary system, so we propose that this modulation represents a surface rotation of $1.299\pm0.001$ days. PG~1159-035 now joins PG~0112+104 \citep{Hermes2017} as one of only two WDs with photometrically detected surface rotation frequencies.  

If  8.906~$\mu$Hz is the surface rotation frequency and PG~1159-035 rotates as a solid body as proposed by
\citet{RePEc:nat:nature:v:461:y:2009:i:7263:d:10.1038_nature08307}, we would expect $\ell$ = 1 triplets with splittings of $\delta\nu\,= 4.4\,\mu$Hz and $\ell = 2$ quintuplets with $\delta\nu\,= 7.4\,\mu$Hz. This predicted $\ell=1$ splitting is at the outer range of permitted values from {\it K2} data.  The predicted $\ell = 2$ splittings are several sigmas from the values observed.

\begin{figure}
    \centering
    \includegraphics[trim= 1.5cm 0 1.3cm 0.5cm, width=0.45\textwidth]{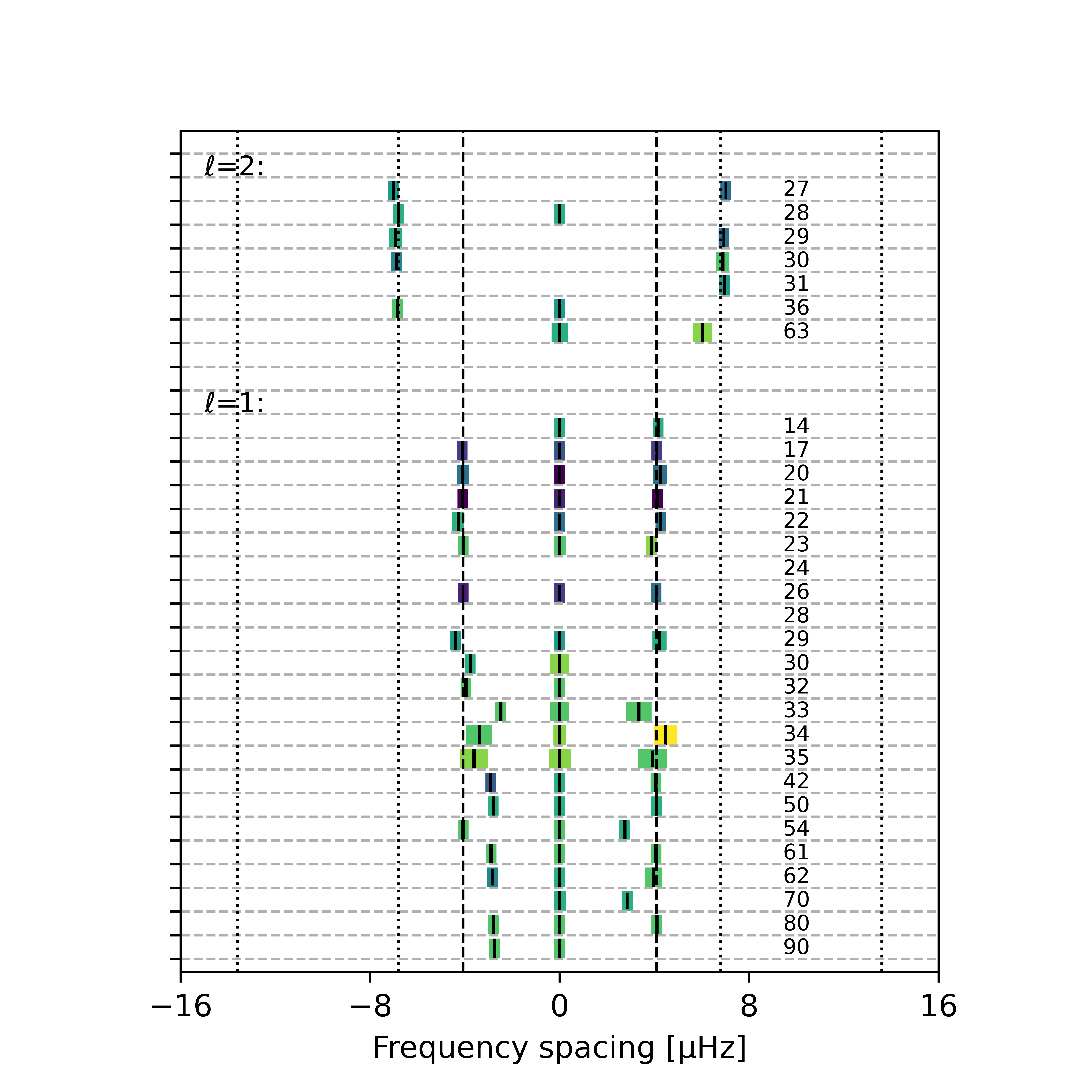}
    \caption{The observed multiplet splittings for 23 $\ell$ = 1 and 9 $\ell$ = 2 modes (we do not plot the singlets) centered on the $m=0$ component.  The $k$ number for each mode is indicated on the right. The dashed and dotted lines indicate the frequency splittings if the star rotates as a rigid body with frequency $\Omega_{\rm rot} = 8.16~\mu \mathrm{Hz}$. The width and color of the boxes correspond respectively to the uncertainty and the amplitude of the frequencies: the darker the color, the higher the amplitude.  }
    \label{rotation}
\end{figure}
Figure~\ref{rotation} shows the observational frequency splitting of $\ell$ = 1 and $\ell$ = 2 compared with the theoretical splitting, if PG~1159-035 rotates as a rigid body with frequency $\Omega_\mathrm{rot} = 8.16\ \mu\mathrm{Hz}$. We do not find good agreement.

\begin{figure}
    \centering
    \includegraphics[width=0.5\textwidth,trim=0.8cm 0cm 0.5cm 1.2cm]{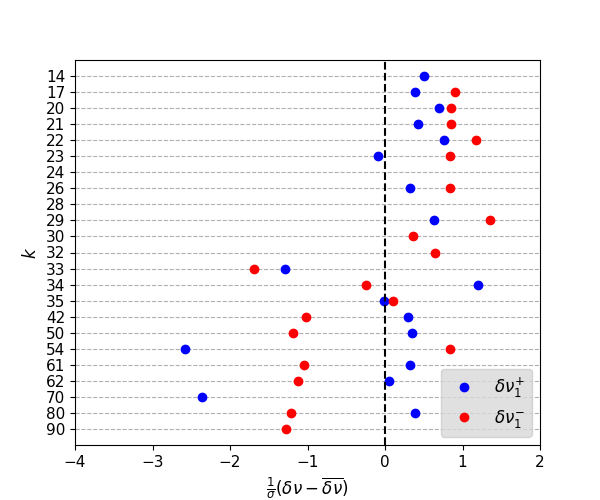}
    \caption{How far each $\delta\nu_1$ is from the median value $\overline{\delta\nu_1}$. It was calculated separately for frequency spacings between $m=-1$ and $m=0$ modes (red points) and for frequency spacings between $m=0$ and $m=1$ modes (blue points).}
    \label{}
\end{figure}

\par The {\it K2} data shows evidence for two different rotation periods, the first from direct detection of a peak in the FT and a second from the average multiplet structure.  If we interpret the rotation period derived from the multiplet structure as a globally averaged rotation rate, this provides evidence that differential rotation plays a role in PG1159-035. Further analysis of differential rotation in PG1159-035 will surely be the subject of future work.  

\begin{figure}[h!]
    \centering
    \includegraphics[width=0.48\textwidth, trim= 0.4cm 0 0  0]{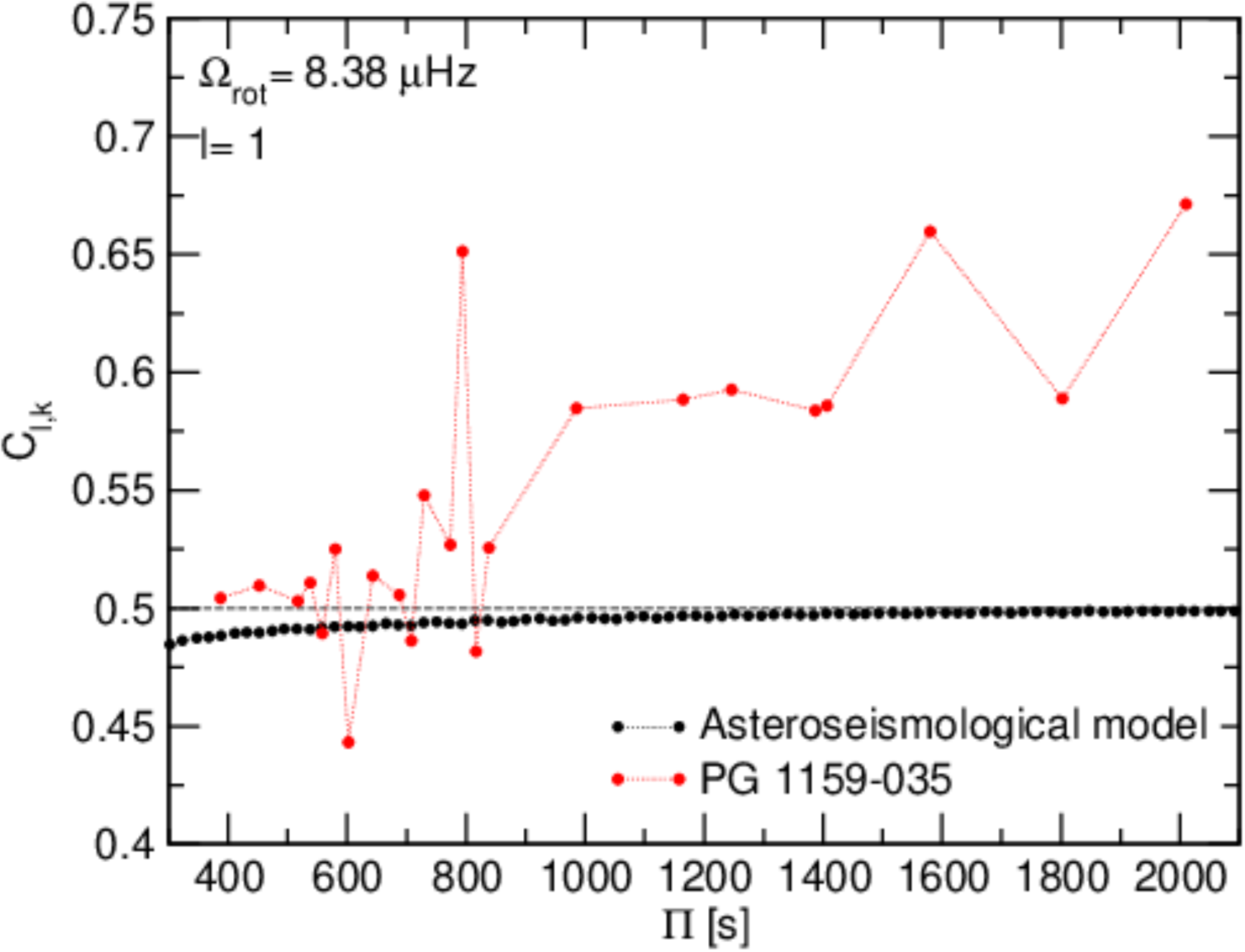}
    \caption{Observed Ledoux rotation coefficients $C(\ell,k)$ (red) versus the predicted ones (black) in the best asteroseismic model, in terms of the periods. The horizontal dotted line corresponding to $C(\ell,k)= 0.5$ is the asymptotic value for large radial-order modes. If $\Omega_\mathrm{rot}=8.38~\mu$Hz, all values are larger than the model ones.}
    \label{clk}
\end{figure}

Figure~\ref{clk} shows the observed Ledoux rotation coefficients $C(\ell,k)$ versus the predicted ones in the best asteroseismic model, which are close to the asymptotic values,
for $\Omega_\mathrm{rot}=8.38~\mu$Hz \citep{Winget91}. The observed values are in general larger than the predicted ones, indicating second order terms are necessary. If we assume the rotation frequency of $\Omega_\mathrm{rot}=8.90~\mu$Hz , all values are larger than the model and asymptotic value.

We observed a significant peak (4.29 mma) in the FT of the {\sl TESS} light curve at 8.66 $\mu$Hz and we used the open source Python package TESS Localize \citep{2022arXiv220406020H} to make sure that  this variability comes from PG1159-035. If this frequency is due the surface rotation, then it can indicate that the rotation frequency of the star not only changes radially, but also temporally.

\section{Period changes}
\label{section9}
\par The evolution of a typical WD is dominated by cooling. An observable effect of WD evolution is change in pulsation periods. Measurements of period change can be used to constrain fundamental physical properties and constrain evolutionary models.  

\par For a typical WD model, we expect the rate of change of period with time  $\dot{\Pi}$ values between $10^{-12}-10^{-15}\, \mathrm{s\,s}^{-1}$.  Hot WDs evolve rapidly, so higher values of $\dot{\Pi}$ will be associated with hotter stars. PG~1159-035 is very hot ($T_\mathrm{eff}\approx 140,000 K$) and is evolving quickly.  The period changes can be measured directly \citep[e.g.][]{2008A&A...489.1225C}.  As PG~1159-035 has not yet completed the contraction of its outer layers, we must also consider the effects of contraction of the stellar atmosphere, which can produce decreasing periods ($\bf \dot{\Pi}<0$).

\begin{figure}[h!]
    \centering
    \includegraphics[width=0.50\textwidth, trim= 1.cm 0.3cm 0  0]{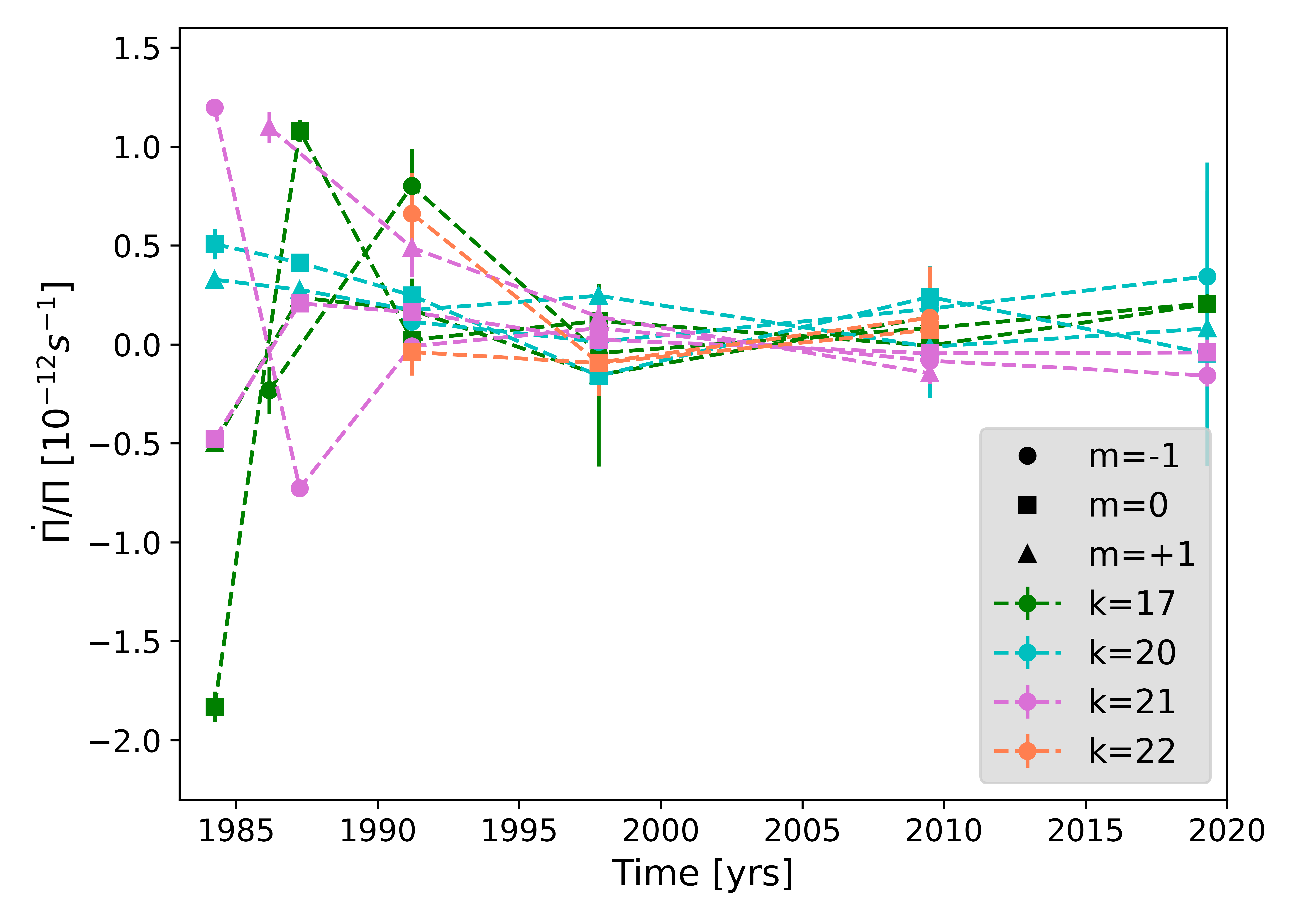}
    \caption{Relative change of pulsation period with time $\frac{\dot{\Pi}}{\Pi}$  over time to $k=17$, $k=20$, $k=21$ and $k=22$ modes. The error bars were calculated by Markov Chain Monte Carlo Simulation.}
    \label{Pdot}
\end{figure}

We used the periods measured in the earlier data cited in Table~\ref{obs_data} to calculate the period changes of PG~1159-035's high-amplitude modes. We choose four modes which often appear in the data. They are the $k=17,20,21,22$ modes, all $\ell=1$. For each given index radial $k$ mode, the period change between two consecutive data sets $n$ and $n+1$ was evaluated as:
\begin{equation}
    \frac{\dot{\Pi}}{\Pi} \left(k, t = \frac{t_{n+1} + t_{n}}{2} \right)= \frac{1}{\Pi_{k}(t_0)} \frac{\Pi_{k}(t_{n+1}) - \Pi_{k}(t_{n})}{t_{n+1} - t_{n}} 
\end{equation}
where $\Pi_{k}(t_{n})$ is the period of index radial $k$ mode observed in the year $t_n$ and $t_0$ is the year of the first observation for the considered datasets. 

\par The period changes of these modes over the years do not present a clear pattern. However, Figure~\ref{Pdot} shows that, overlapping, they look to be converging to some value and then scattering again. The {\it m} components of the $\it k$=17 and the $\it k$=21 modes switch between positive and negative $\dot{\Pi}$ values.  Clearly, we are neglecting an important effect that acts on timescales of months and years, as nonlinear mode coupling. Other possible effects include reflex motion and proper motion, but these would be expected to act equally on all modes.  Magnetic fields would also affect $\dot{\Pi}$, and could affect each mode differently. 

\section{Combination frequencies}
\label{section10}

\par It was previously reported that PG~1159-035 had no 
combination frequencies, since none were detected in the previous data \citep[e.g.][]{Costa08}. However, due to the extended time base and signal-to-noise of the \emph{K2} data set, for the first time we are able to identify combination frequencies in its pulsation spectrum (see Table~\ref{tab:lincom}).

The amplitudes of these frequencies are small, but then so are those of the parent linear modes. A measure of the strength of the nonlinearity is
\begin{equation}
    R_{\rm c} \equiv \frac{A_{i+j}}{n_{ij} A_i A_j}
\end{equation}
where $A_i$ and $A_j$ are the parent amplitudes (in units of modulation amplitude ma), $A_{i+j}$ is the amplitude of the combination frequency, and $n_{ij} =2$ for $i \neq j$ and 1 otherwise \citep{vanKerkwijk00,Yeates05}.

From Table~\ref{tab:lincom} we find $R_{\rm c}$ values in the range $\approx 4$--14. We note that these values are not smaller than those found in the DAVs, and are in fact completely consistent with them \citep[see][]{Yeates05}.

In the DAVs and DBVs, the dominant mechanism producing combination frequencies and nonlinear light curves is thought to be the interaction of the pulsations with the surface convection zone \citep{Brickhill92a,Brickhill92b,Wu01,Montgomery05a}.
PG~1159-035 is sufficiently hot that models show no surface convection zone \citep[e.g.,][]{Werner89}, so the presence of combination frequencies is something of a mystery, though \citet{Kurtz15} find multiple large apparent amplitude combination peaks in Slowly Pulsating B stars and $\gamma$~Dor stars, and explain that visibility, which is highly dependent on the surface patterns of oscillations, can explain these amplitudes apparently larger than their parent frequencies. 

An alternative mechanism is the nonlinear temperature to flux relationship described in \citet{Brassard95}. While this mechanism is not generally successful in explaining the nonlinearities seen in the DAVs \citep[e.g., ][]{Vuille00}, we consider whether it could be viable for the amplitudes of the combination frequencies observed in the DOV\footnote{Pulsating PG~1159 stars.} PG~1159-035. 

To simplify the calculation, we approximate the stellar flux as a blackbody, $B_\lambda(T)$, and expand it to second order in the temperature perturbations, $\Delta T$:
\begin{equation}
    F_\lambda \approx B_\lambda(T_0) +
    \frac{\partial B_\lambda(T_0)}{\partial T} \Delta T +
    \frac{1}{2}\frac{\partial^2 B_\lambda(T_0)}{\partial T^2} (\Delta T)^2 + \ldots
\end{equation}
Assuming just one mode with temperature perturbation $\Delta T_i$, we find that the fractional flux perturbations are
\begin{eqnarray}
    A_i & = & \frac{1}{B_\lambda(T_0)} \frac{\partial B_\lambda(T_0)}{\partial T} \Delta T_i \\
    A_{2i} & = & \frac{1}{2 B_\lambda(T_0)} \frac{\partial^2 B_\lambda(T_0)}{\partial T^2} \Delta T_i^2
\end{eqnarray}
with the result that
\begin{equation}
    R_{\rm c} = \frac{A_{2i}}{A_i^2} = \frac{1}{2} B_\lambda(T_0) \left( \frac{\partial B_\lambda(T_0)}{\partial T} \right)^{-2}
    \frac{\partial^2 B_\lambda(T_0)}{\partial T^2} 
     \label{eqRc}
\end{equation}

\begin{figure}
    \centering
    \includegraphics[width=0.45\textwidth,trim=0.8cm 0 0 0]{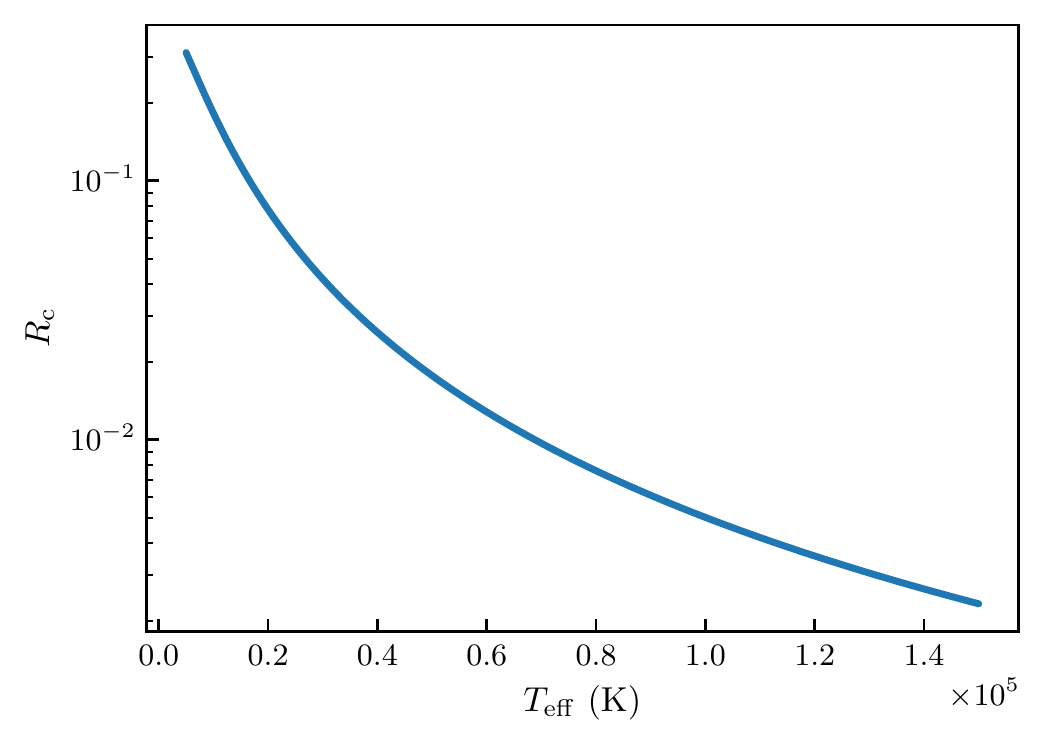}
    \caption{Expected value of $R_{\rm c}$ (Equation~\ref{eqRc}) as a function of $T_{\rm eff}$ for observations centered on $\lambda = 5500\,$\AA. A blackbody spectrum was assumed for this calculation.}
    \label{R_c}
\end{figure}

\begin{table*}[t!]
    \centering

    \begin{tabular}{|ccccc|}\hline
      Frequency ($\mu$Hz)           & Amplitude (mma)        & Combination frequencies $f_k$  & Difference ($\mu$Hz) & $R_{\rm c}$ \\ \hline \hline

      $75.31 \pm 0.03$    & $0.17 \pm  0.05$  & $f_{20} - f_{21}$   & $-$0.06   (2.1$\,\sigma$)  & 3.9\\
      $142.74 \pm 0.03$   & $0.19 \pm  0.05$  & $f_{20} - f_{22}$   & +0.01   (0.5$\,\sigma$) & 11.6\\
      $419.28 \pm 0.06$   & $0.08 \pm  0.05$  & $f_{17} - f_{22}$   & +0.25   (4.2$\,\sigma$) & 14.0\\
      $3412.82 \pm 0.03$  & $0.20 \pm  0.05$  & $f_{21} + f_{26}$   & +0.01  (0.1$\,\sigma$) & 9.6 \\
      $3716.39 \pm 0.02$  & $0.24 \pm  0.05$  & $2 f_{21}$          & $-$0.01  (0.5$\,\sigma$) & 14.4\\
      $3791.56 \pm 0.03$  & $0.18 \pm  0.05$  & $f_{20} + f_{21}$   & +0.06  (1.7$\,\sigma$) & 4.1\\
      $3866.84 \pm 0.03$  & $0.19 \pm  0.05$  & $2 f_{20}$          & +0.03  (0.7$\,\sigma$) & 6.7\\
      $4143.67 \pm 0.03$  & $0.15 \pm  0.05$  & $f_{17} + f_{20}$   & $-$0.03  (1.0$\,\sigma$) & 7.5\\ \hline
    \end{tabular}

    \caption{Combination frequencies in {\em K2} data.}
    \label{tab:lincom}
\end{table*} 

In Figure~\ref{R_c}, we plot $R_{\rm c}$ as a function of $T$ for observations centered at the wavelength $\lambda = 5500\,$\AA.
At $T_{\rm eff} \approx\,$140,000~K, we see that $R_{\rm c} < 0.01$, which is much smaller than the observed values. Thus, if a blackbody spectrum is a good proxy for the actual flux distribution, then the mechanism of \citet{Brassard95} cannot explain the observed nonlinearities in PG1159-035. However, it is possible that using actual model atmospheres could produce larger values of $R_{\rm c}$, 
but the atmospheres would need to be calculated on a fine enough grid in $T_{\rm eff}$ that accurate first- and second-order derivatives can be computed.
Clearly, understanding the origin of these combination frequencies will require further analysis. 

\section{Asteroseismic modeling}
\label{astero-model}

\begin{figure}
    \centering
    \includegraphics[width=0.46\textwidth,trim=0.6cm 0 0 0]{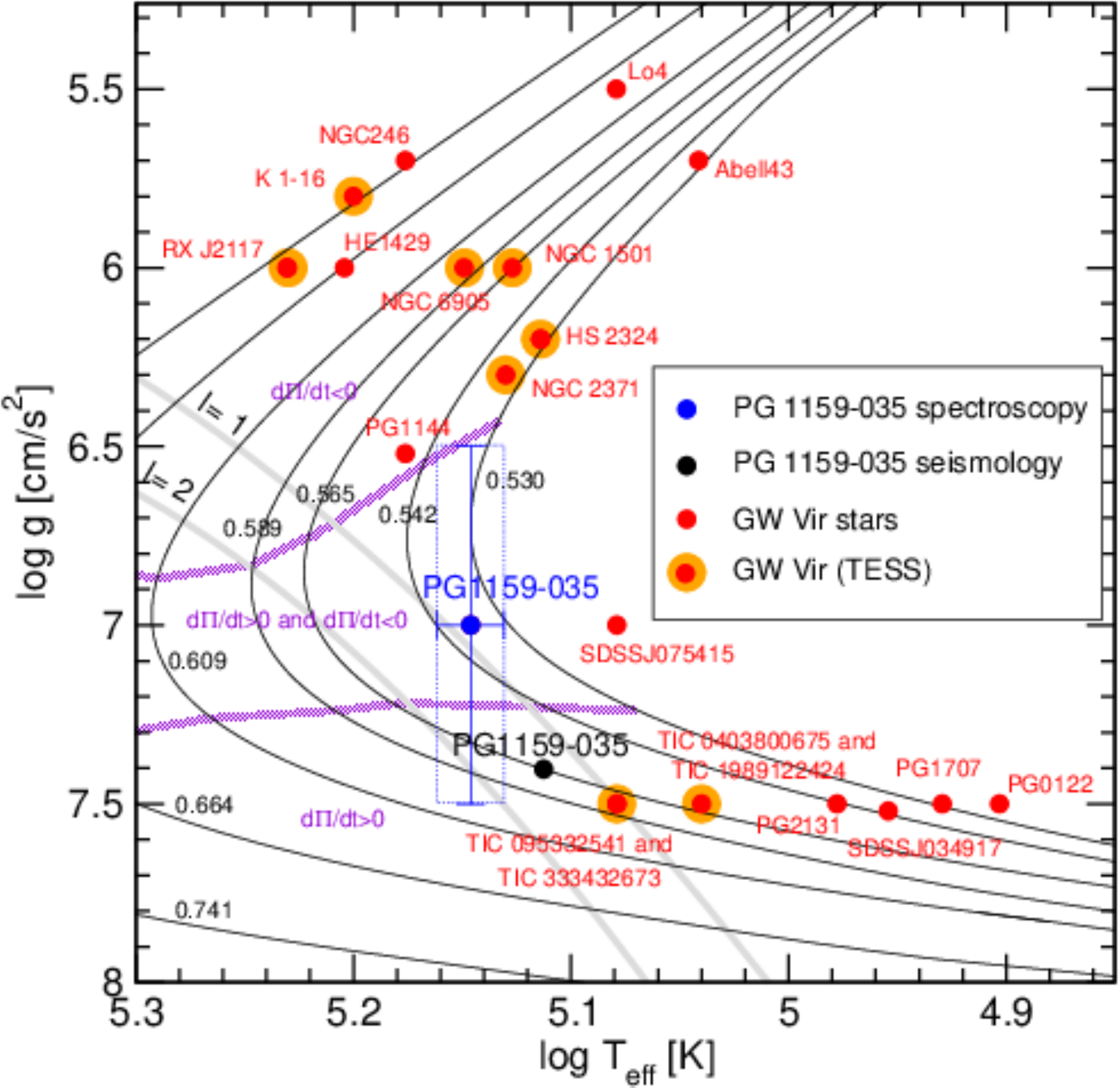}
\caption{Location of the known GW~Vir variable stars
  in the  $\log T_{\rm eff}- \log g$
  diagram, depicted with small red circles. Thin   solid   curves   show   the
  post-born-again   evolutionary tracks from  \citet{2005A&A...435..631A} and \citet{2006A&A...454..845M}
  for  different  stellar  masses in the range $0.530-0.741 M_{\odot}$. 
  The location of the GW~Vir stars
  observed  with {\sl TESS} (with published results) 
  are emphasized with large orange circles. PG~1159$-$035 is displayed with a small blue circle  
  with the error bars (within a box with dotted lines), according to \citet{2011A&A...531A.146W} ($T_{\rm eff}= 140\,000\pm5\,000$ K and $\log g= 7 \pm 0.5$). The location of PG~1159$-$035 according to the asteroseismic 
  model (see Sect. \ref{period_fits}) is depicted with a black, small circle.
  The violet curves divide the plane in three regions: one in which all the $g$ modes have $\dot{\Pi}<0$, another region where some modes have $\dot{\Pi}<0$ and others have $\dot{\Pi}>0$, and finally a region in which all the modes 
  have $\dot{\Pi}>0$. The gray curves correspond to the theoretical blue edge of the GW Vir instability strip for $\ell= 1$ and $\ell= 2$ 
  $g$ modes, according to \citet{2006A&A...458..259C}.
\label{logg-teff}
}
\end{figure}

The location of PG~1159-035 in the $\log T_{\rm eff}- \log g$ plane is displayed in Fig. \ref{logg-teff} with a blue dot with error bars. If the 
star has $T_{\rm eff}= 140\,000\pm5\,000$ K and $\log g= 7.0 \pm 0.5$ \citep{2011A&A...531A.146W}, the PG~1159 evolutionary tracks of \citet{2005A&A...435..631A} and \citet{2006A&A...454..845M} 
indicate that the star has just turned its ``evolutionary knee'' 
(maximum $T_{\rm eff}$), and is entering its WD cooling track.  
The spectroscopic stellar mass of the star, considering 
the uncertainties in $T_{\rm eff}$ and $\log g$, is derived by linear interpolation 
and results in $M_{\star}= 0.54\pm0.07 M_{\odot}$. We note that 
the star falls in the region where our pulsation models predict 
a mix of positive and negative rates of period changes (Fig. \ref{logg-teff}), 
as found by \citet{2008A&A...489.1225C} 
(see also Sect. \ref{section8}). 

In the next sections, we first describe the PG~1159 evolutionary models used in this work, and then we apply the tools of WD asteroseismology for inferring the stellar mass and the derivation of an asteroseismic model for PG~1159$-$035. Sect. \ref{tess-data} 
shows that {\sl TESS} has detected very few periods (see Table \ref{tab:tess}), and they are almost identical to the corresponding \emph{K2} periods. Therefore, we will carry out our asteroseismic modeling considering only the \emph{K2} periods (Tables \ref{tab:l=1} and \ref{tab:l=2}). 

\subsection{PG~1159 stellar models}
\label{models}

The PG1159 star model set used in this work has been described in depth in \citet{2005A&A...435..631A} and \citet{2006A&A...454..845M,2007A&A...470..675M}. We refer the interested reader to those papers for details. \cite{2005A&A...435..631A}
and \cite{2006A&A...454..845M} computed the complete evolution
of non-rotating model star sequences with initial masses on  the ZAMS in the range $1 - 3.75\ M_{\sun}$  and assuming  a metallicity of $Z= 0.02$.  All the  post-AGB evolutionary sequences  computed with the {\tt
  LPCODE} evolutionary code \citep{2005A&A...435..631A} were followed
through  the  very  late   thermal  pulse  (VLTP)  and  the  resulting
born-again  episode that  give rise  to the  H-deficient, He-,  C- and
O-rich composition characteristic of  PG~1159 stars.  The masses of the
resulting  remnants are  $0.530$, $0.542$, $0.565$, $0.589$, $0.609$,
$0.664$, and  $0.741 \ M_{\sun}$.  In Fig.~\ref{logg-teff} the
evolutionary tracks employed  in this  work are shown in the $\log
T_{\rm eff}$ vs. $\log g$ plane.

\subsection{Derivation of the stellar mass from the period spacings}
\label{mass-period-spacing}

\begin{figure}[ht]
    \centering
    \includegraphics[width = 0.46\textwidth, trim=0.5cm 0 0 0]{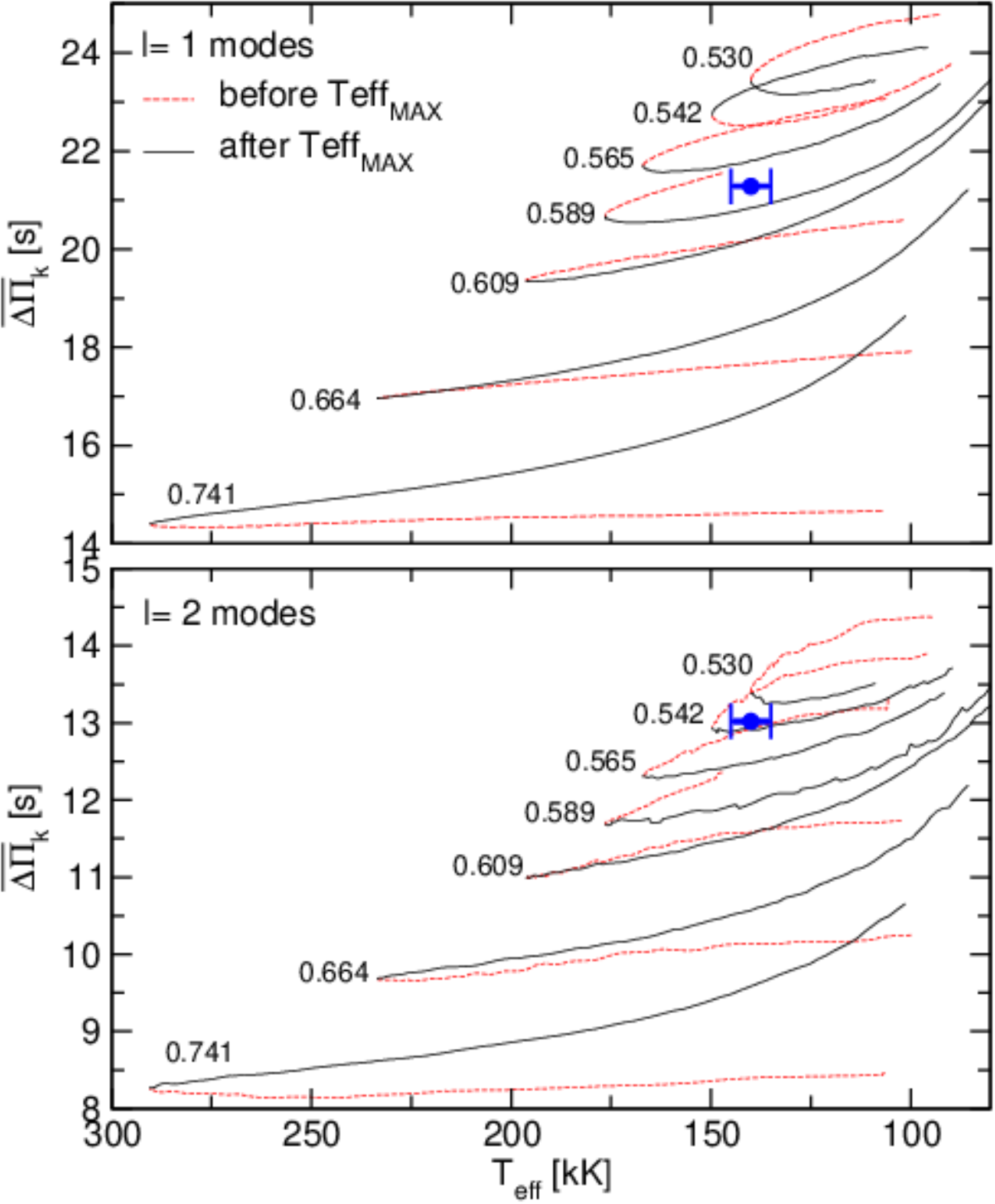}
\caption{{\sl Upper panel}: dipole ($\ell= 1$) average of the computed period spacings, $\overline{\Delta \Pi_k}$, assessed  in  a  range  of  periods  that
  includes  the  periods  observed  in PG~1159$-$035, shown as dashed red
  (solid black) curves corresponding to stages before (after) the maximum
  $T_{\rm eff}$ for different stellar masses. The location of PG~1159$-$035
  when we  use the effective temperature derived by
  \citet{2011A&A...531A.146W}, $T_{\rm eff}= 140\,000 \pm 5\,000$~K,
  and the period spacing $\Delta \Pi_1= 21.28\pm0.02$~s derived in
  Sect.~\ref{section6}, is highlighted with a blue circle. {\sl Lower panel}: 
  same as in upper panel, but for the case in which the period spacing is assumed to be associated to $\ell= 2$ modes ($\Delta \Pi_2= 13.02\pm0.04$~s).}
\label{psp}
\end{figure}

The approach we use to extract information on the stellar
mass of PG~1159$-$035 is the same employed in, e.g., \citet{2021A&A...645A.117C}. 
Briefly, a way to estimate
the masses of GW~Vir stars is by comparing the
observed period spacing $\Delta \Pi$ with the
asymptotic period spacing $\Delta \Pi_{\ell}^{\rm a}$ \citep[][]{1990ApJS...72..335T}
at the effective temperature of the star, following the pioneering work of \citet{1988IAUS..123..329K}. Since GW Vir stars generally do not have 
all of their pulsation modes in the asymptotic regime, and  
are not chemically homogeneous, it is more reliable
to infer the stellar mass by comparing $\Delta \Pi$ with
the average of the computed period spacings ($\overline{\Delta
  \Pi_{k}}$). It is assessed  as $\overline{\Delta \Pi_{k}}= (n_{\rm c}-1)^{-1} \sum_{k=1}^{n_{\rm c}} \Delta \Pi_{k}$, where the ``forward'' period spacing ($\Delta \Pi_{k}$) is defined as $\Delta \Pi_{k}= \Pi_{k+1}-\Pi_{k}$ ($k$ being the radial order) and $n_{\rm c}$ is the number of computed periods laying in the range of the observed periods.   Note that this method for assessing the stellar mass relies on the spectroscopic effective temperature, and the results are unavoidably affected by its associated uncertainty.

We have calculated the average of the computed period spacings for
$\ell= 1$ and $\ell= 2$, in terms of the effective
temperature, for all the masses considered. We employed the 
LP-PUL pulsation code \citep{2006A&A...454..863C} for computing the dipole and quadrupole adiabatic periods of $g$ modes on fully evolutionary PG1159 models generated with 
the LPCODE evolutionary code \citep{2005A&A...435..631A}. 
The results  are shown in
the upper ($\ell= 1$) and lower ($\ell= 2$) panels of Fig.~\ref{psp}, where we show  $\overline{\Delta
  \Pi_{k}}$ corresponding to evolutionary stages before the maximum possible  effective temperature, ${T_{\rm eff}}^{\rm MAX}$, which depends on the  stellar mass, with red dashed lines, and the  phases after that ${T_{\rm eff}}^{\rm MAX}$, the so-called WD stage itself,  with solid black lines. The location of PG~1159$-$035 is indicated by a small blue circle with error bars, and corresponds to the effective temperature of the star  according to \citet{2011A&A...531A.146W} and the period spacings derived in Sect.~\ref{section6}. We perform linear interpolations between the sequences and obtain the stellar mass shown in Table \ref{spacing}. If the star is after the ``evolutionary knee'', as suggested by its spectroscopic parameters (see Fig. \ref{logg-teff}), then the stellar mass is $\sim 0.58 M_{\odot}$ according to the  $\ell= 1$ modes, and $\sim 0.54 M_{\odot}$ according to the $\ell= 2$ modes. 

\begin{table}[ht]
\centering
\begin{tabular}{l|cc}
\hline
\hline
& $\ell= 1$ & $\ell= 2$ \\
\hline
Before the maximum $T_{\rm eff}$  & $0.594^{+0.003}_{-0.002}$ & $0.560^{+0.005}_{-0.018}$ \\
                                  &                           &                           \\ 
After the maximum $T_{\rm eff}$   & $0.576^{+0.005}_{-0.004}$ & $0.538^{+0.004}_{-0.002}$ \\
\hline
\end{tabular}
\caption{Stellar mass (in $M_{\odot}$) derived for  
PG~1159$-$035 by comparing the the average of  the computed period
spacings  ($\overline{\Delta \Pi_k}$)  of  our PG~1159 models  with  the observed period spacings derived in Sect.~\ref{section6}.}
\label{spacing}
\end{table} 

If the star were at an earlier stage, before the``evolutionary knee'', the mass would be $\sim 0.59 M_{\odot}$ and $\sim 0.54 M_{\odot}$, according to the modes with $\ell= 1$ and $\ell=2$, respectively. We conclude that the stellar mass of PG~1159$-$035, as derived from its period spacings $\Delta \Pi_1$ and $\Delta \Pi_2$, is in the range $0.54-0.59 M_{\odot}$. This range of masses is compatible with the spectroscopic mass, 
$M_{\star}= 0.54 \pm 0.07\, M_{\odot}$ \citep{2011A&A...531A.146W}. 

\subsection{Asteroseismic period fits}
\label{period_fits}

\begin{figure*}[ht!]
\includegraphics[clip,width= 1.0\textwidth]{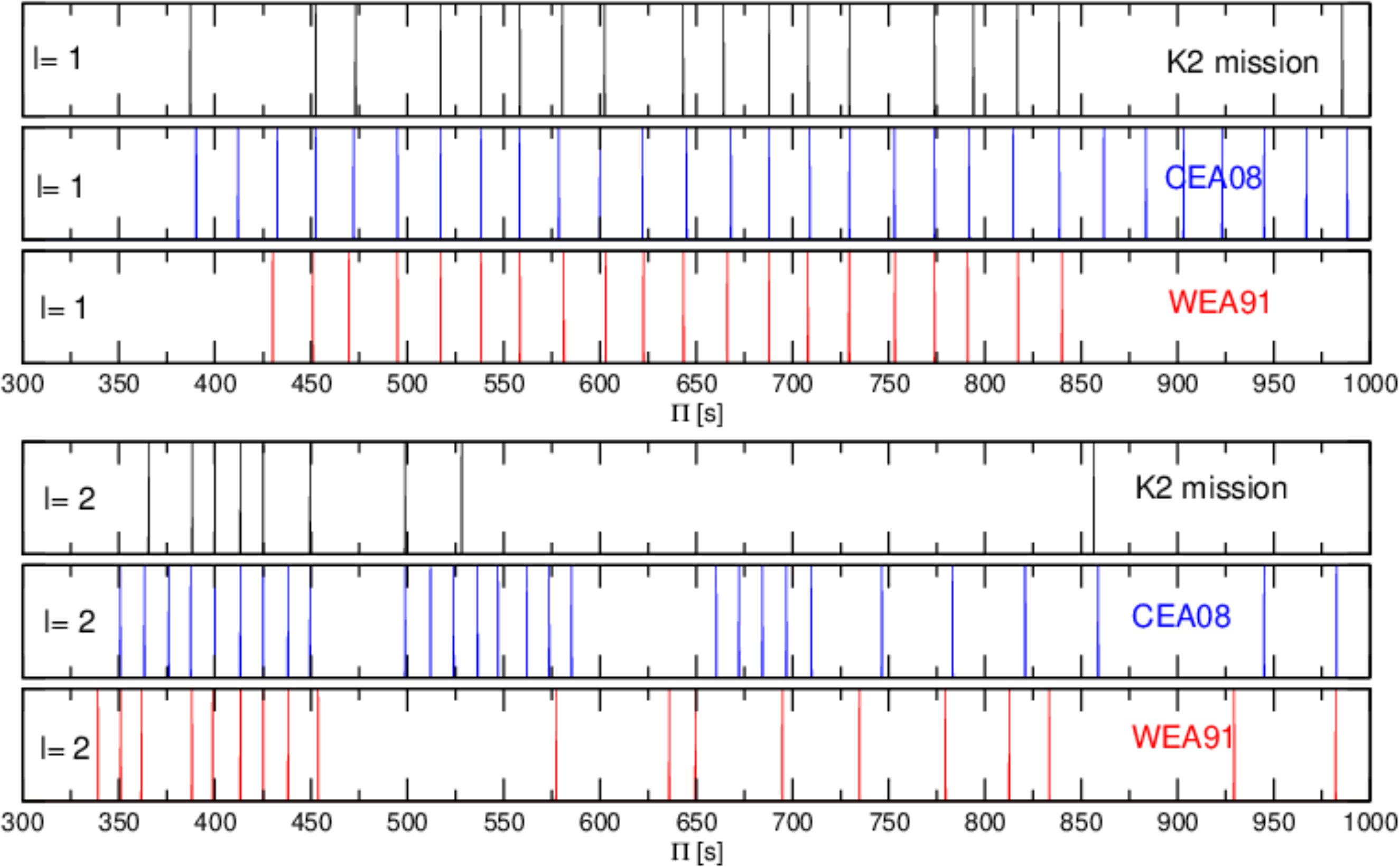}
\caption{Schematic distribution of the $\ell= 1$ and $\ell= 2$ ($m= 0$) pulsation periods of PG~1159$-$035
  observed by {\sl K2} (black lines, upper panels), 
  observed by \citet{Costa08} (CEA08, blue lines,
  middle panels), and observed by \citet{Winget91} (WEA91, red lines, lower panels). At least 14 additional $\ell =1$ modes only detected by 
  {\sl K2} at periods longer than 1000 s have been cut off and not shown here, but are detailed in Table \ref{tab:l=1}. The amplitudes have been arbitrarily  set to one for
  clarity.}
\label{compara-m0} 
\end{figure*} 

An asteroseismic tool to disentangle the internal structure
of GW~Vir stars is to search for theoretical models that best match the
individual pulsation  periods  of  the target star. To measure the
goodness of the agreement between the theoretical periods
($\Pi_{\ell,k}$) and the observed periods ($\Pi_i^{\rm
  o}$), we adopt the quality function: $\chi^2(M_{\star}, T_{\rm eff})= \frac{1}{N} \sum_{i= 1}^{N} {\rm min}[(\Pi_{\ell,k}-\Pi_i^{\rm o})^2]$ \citep{2021A&A...645A.117C},
  where $N$ is the number of observed periods. In order to find the
stellar model that best replicates the observed periods exhibited by
each target star -- the ``asteroseismic'' model --, we
assess the  function  $\chi^2$ for stellar model  masses  $M_{\star}=
0.530, 0.542,  0.565, 0.589, 0.609, 0.664$, and $0.741 M_{\odot}$. For
the  effective  temperature, we  employ  a much  finer  grid ($\Delta
T_{\rm eff}= 10-30$ K).  The PG~1159 model that
shows the lowest value of $\chi^2$ is adopted as the best-fit
asteroseismic model. 

\begin{table*}[ht]
\centering
\begin{tabular}{cc|cccccl}
\hline
\noalign{\smallskip}
$\Pi_i^{\rm O}$ & $\ell^{\rm O}$ & $\Pi_k$ & $\ell$ & $k$ &   $\delta \Pi_k$ & $\dot{\Pi}_k$ & 
Unstable\\ 
(s) & & (s) &  & & (s) & ($10^{-11}$ s/s) &  \\
\noalign{\smallskip}
\hline
\noalign{\smallskip}       
 (387.19)     & 1 &  388.29 & 1 & 16 & $-1.10$ & 1.11 & no \\
 452.45       & 1 &  452.46 & 1 & 19 & $-0.01$ & 1.15 & no \\  
 473.06       & 1 &  474.24 & 1 & 20 & $-1.18$ & 0.81 & no \\ 
 517.22       & 1 &  515.69 & 1 & 22 &  1.53 & 1.22 & no \\
 538.16       & 1 &  537.78 & 1 & 23 &  0.38 & 1.07 & no \\
 558.45       & 1 &  557.60 & 1 & 24 &  0.85 & 0.60 & no \\
 580.40       & 1 &  579.02 & 1 & 25 &  1.38 & 1.29 & no \\
 (602.35)     & 1 &  601.90 & 1 & 26 &  0.45 & 1.22 & no \\
 643.24       & 1 &  642.93 & 1 & 28 &  0.31 & 1.51 & no \\
 664.20       & 1 &  665.31 & 1 & 29 & $-1.11$ & 1.07 & no \\
 687.74       & 1 &  685.85 & 1 & 30 &  1.89 & 1.01 & no \\
 708.12       & 1 &  707.25 & 1 & 31 &  0.87 & 1.60 & no \\
 729.58       & 1 &  728.55 & 1 & 32 &  1.03 & 1.30 & no \\
 773.71       & 1 &  771.82 & 1 & 34 &  1.89 & 1.49 & no \\
 793.95       & 1 &  793.26 & 1 & 35 &  0.69 & 1.85 & no \\
 816.74       & 1 &  814.77 & 1 & 36 &  1.97 & 1.31 & no \\
 838.36       & 1 &  836.14 & 1 & 37 &  2.22 & 1.66 & no \\
 985.65       & 1 &  987.59 & 1 & 44 & $-1.94$ & 1.87 & no \\
1116.01       & 1 & 1120.53 & 1 & 50 & $-4.52$ & 2.62 & no \\
1164.88       & 1 & 1163.72 & 1 & 52 &  1.16 & 2.95 & no \\
1246.27       & 1 & 1251.47 & 1 & 56 & $-5.20$ & 1.95 & no \\
1284.52       & 1 & 1273.97 & 1 & 57 & 10.55 & 3.32 & no \\
1387.06       & 1 & 1383.81 & 1 & 62 &  3.25 & 3.62 & no \\
1406.51       & 1 & 1405.39 & 1 & 63 &  1.12 & 2.37 & no \\
1539.03       & 1 & 1538.67 & 1 & 69 &  0.36 & 3.45 & no \\
1555.74       & 1 & 1560.10 & 1 & 70 & $-4.36$ & 3.27 & no \\
1580.08       & 1 & 1581.27 & 1 & 71 & $-1.19$ & 3.58 & no \\
1802.11       & 1 & 1802.75 & 1 & 81 & $-0.64$ & 4.03 & no \\
1982.78       & 1 & 1981.36 & 1 & 89 &  1.42 & 5.18 & no \\
2010.07       & 1 & 2002.26 & 1 & 90 &  7.81 & 3.15 & no \\
2084.71       & 1 & 2091.45 & 1 & 94 & $-6.74$ & 6.57 & no \\
2807.93       & 1 & 2804.11 & 1 &126 &  3.82 & 9.17 & no \\
\hline
365.54        & 2 &  364.02 & 2 & 27 &  1.52 & 0.70 & no \\
(388.18)      & 2 &  388.36 & 2 & 29 & $-0.18$ & 0.51 & no \\
400.08        & 2 &  400.67 & 2 & 30 & $-0.59$ & 0.88 & no \\
(413.22)      & 2 &  413.73 & 2 & 31 & $-0.51$ & 0.93 & no \\
(425.00)      & 2 &  425.42 & 2 & 32 & $-0.42$ & 0.72 & no \\
449.36        & 2 &  450.18 & 2 & 34 & $-0.82$ & 1.18 & no \\
498.74        & 2 &  500.55 & 2 & 38 & $-1.81$ & 1.28 & no \\
528.21        & 2 &  524.37 & 2 & 40 &  3.84 & 1.10 & no \\
856.56        & 2 &  852.23 & 2 & 66 &  4.33 & 1.62 & no \\
\noalign{\smallskip}
\hline
\end{tabular}
\caption{Observed $m= 0$ and theoretical periods of the asteroseismic
  model for PG~1159$-$035 [$M_{\star}= 0.565 M_{\odot}$, $T_{\rm eff}=
    129\,600$ K, $\log(L_{\star}/L_{\odot})= 2.189$]. Periods are in
  seconds  and rates of period change  (theoretical) are  in units of
  $10^{-11}$ s/s. $\delta \Pi_i= \Pi^{\rm O}_i-\Pi_k$ represents  the
  period differences, the model $\ell$ the harmonic degree, $k$ the radial
  order, $m$ the azimuthal index.  The last column provides information
  about the pulsational stability/instability   nature  of  the
  modes. Parenthesis indicate $m=0$ periods which are actually 
  absent from the power spectrum,  their values being estimated 
  by averaging the components $m=\pm1$. 
  }
\label{tab:per-l1l2m0}
\end{table*}

We employ the periods identified with $\ell= 1$ and $\ell= 2$ modes
of Tables \ref{tab:l=1} and \ref{tab:l=2}, respectively. 
We consider only $m= 0$ components in the case of multiplets. When frequency multiplets  have the central component ($m= 0$) absent, we adopt a value for this 
frequency, estimating it as the mean value between the components $m= -1$ and $m= +1$ (if both frequencies exist). We have 32 $\ell= 1$ periods and 9 
$\ell= 2$ periods available to perform the period fits. These periods are shown in Table \ref{tab:per-l1l2m0}, and they are also plotted in Fig. \ref{compara-m0}, along with the $m= 0$ periods of \citet{Winget91} (WEA91) and \citet{Costa08} (CEA08). Regarding $\ell= 1$ modes
(the 3 upper panels in Fig. \ref{compara-m0}), we can observe that, in general, the periods which are common to the three data sets are in excellent agreement with each other. This is encouraging, because the results of the previous works, obtained from extensive ground-based observations, are confirmed with the new space data. Second, we can notice that for periods shorter than $\sim 1000$ s, the {\sl K2} observations have fewer periods than those of WEA91 and CEA08. 
Finally, we draw attention to the fact that in the {\sl K2} data there are many periods longer than $\sim 1000$ s, which are not present in the sets derived from ground-based observations. These are new periods of PG~1159$-$035. These long periods were not detectable by ground-based observations, mainly because of the relatively short length of each 
data set and the variable effects of extinction due to the Earth's atmosphere. These effects generate frequency dependent noise, limiting sensitivity to longer periods. As for $\ell= 2$ modes (the 3 lower panels in Fig. \ref{compara-m0}), we note again much fewer periods in the {\sl K2} observations compared to the other two data sets. On the other hand, there is good agreement between the {\sl K2} periods and those of CEA08. We also note, in passing, that the periods from WEA91 longer than $\sim 500$ s are somewhat different from those of CEA08, possibly due to alias contamination in the older data sets. 

\begin{table}
\centering
\begin{tabular}{l|cc}
\hline
\hline
Quantity & Spectroscopy &  Asteroseismology \\
         & Astrometry   &   (This work)      \\ 
\hline
$T_{\rm eff}$ [K]                            & $140\,000 \pm 5\,000^{\rm (a)}$     & $129\,600\pm 2\,000$ \\
$M_{\star}$ [$M_{\odot}$]                     & $0.54 \pm 0.07$           & $0.565\pm 0.008$ \\ 
$\log g$ [cm/s$^2$]                           & $7.0\pm0.5^{\rm (a)}$     & $7.41\pm 0.11$  \\ 
$\log (L_{\star}/L_{\odot})$                  & $2.58 \pm 0.29^{\rm (a)}$ & $2.19\pm 0.04$ \\  
$\log(R_{\star}/R_{\odot})$                   & $\cdots$                  & $-1.61\pm0.05$ \\  
$M_{\rm env}$ [$M_{\odot}$]                   & $\cdots$                  & $0.017$ \\  
$(X_{\rm He},X_{\rm C},X_{\rm O})_{\rm s}$ &   0.33,  0.48,  0.17$^{\rm (a)}$  & 0.386, 0.321,  0.217  \\   
$d$  [pc]                                     & $592\pm 21^{\rm (b)}$  & $444^{+69}_{-59}$  \\ 
$\pi$ [mas]                                   & $1.69\pm0.06^{\rm (b)}$     & $2.25^{+0.35}_{-0.30}$\\ 
$A_{\rm V}$                                   & $\cdots$                    &     $0.064^{+0.002}_{-0.001}$               \\
\hline
\hline
\end{tabular}
\caption{The main characteristics of the GW Vir star PG~1159$-$035. 
The second column  corresponds to spectroscopic and astrometric results, whereas the third column present results from the asteroseismic model of this work.} 
\label{table:modelo-sismo}

{\footnotesize  References: (a)  \citet{2011A&A...531A.146W};  (b) {\sl Gaia} ({\tt https://gea.esac.esa.int/archive/}).}
\end{table}

\begin{figure*} [t]
\includegraphics[clip,width= 1.0\textwidth]{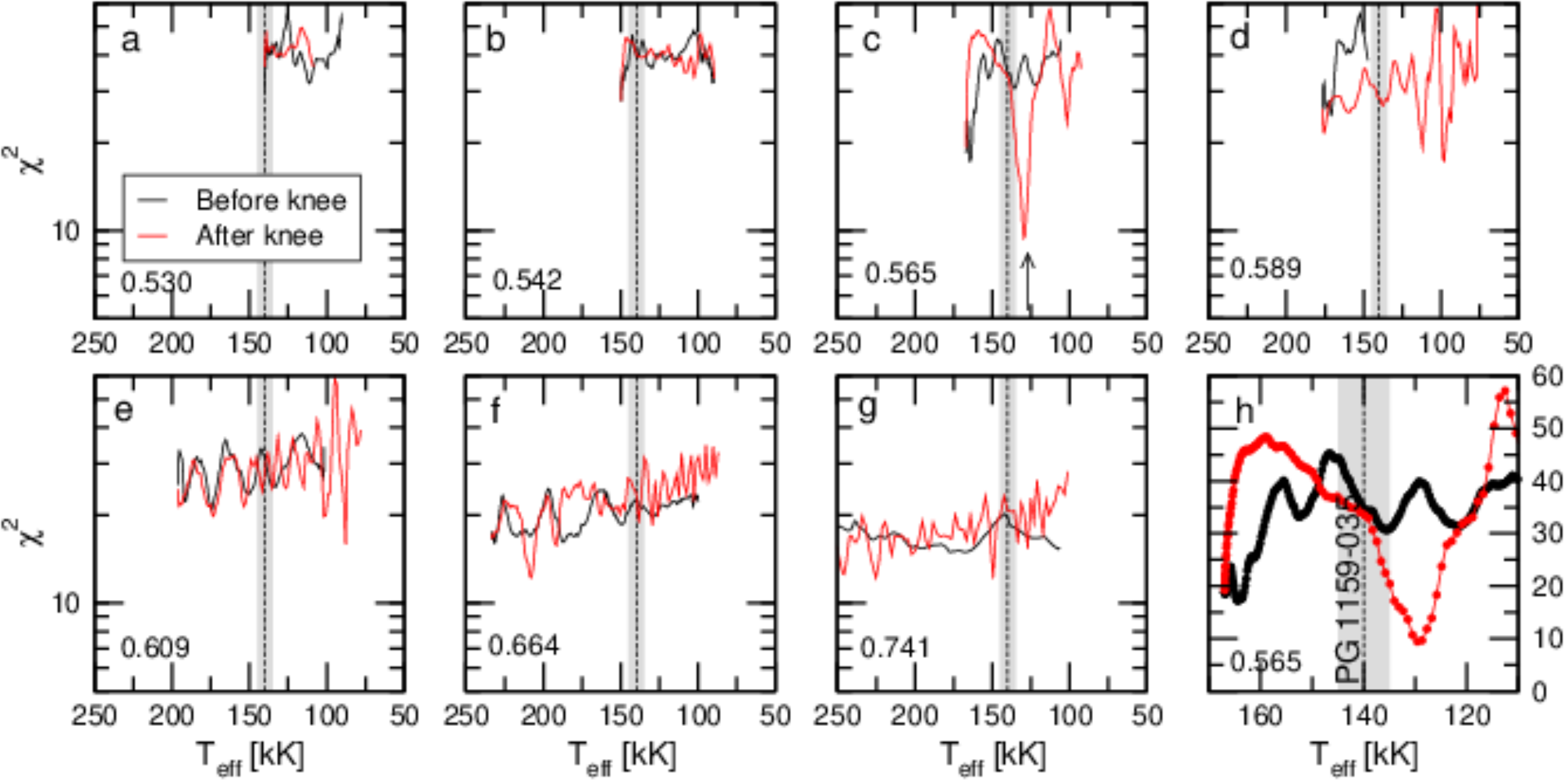}
\caption{The quality  function  of the  period  fit in terms of  the
  effective temperature  for the PG~1159 sequences with different
  stellar masses  (in solar units), indicated  at  the  left-bottom
  corner  of  each  panel.   Black (red) lines correspond to stages
  before (after) the "evolutionary knee" (see Fig. \ref{logg-teff}).   Only
  the periods with $m= 0$ (see Tables \ref{tab:l=1}  and \ref{tab:l=2})
  have been considered. There is a  strong minimum  in  panel  c 
  (marked with an arrow),
  corresponding  to a model with $M_{\star}= 0.565 M_{\odot}$ and 
  $T_{\rm eff}\simeq  129\,600$ K.  
  Panel h is a
  zoom of  the region with the  strong minimum seen in panel c;
  the $y$-axis scale is linear in this case. The vertical dashed line
  is the spectroscopic  $T_{\rm eff}$ of  PG~1159$-$035  (140~kK)  and  the
  gray zone depicts its 1$\sigma$ uncertainty ($\pm 5$~kK).}
\label{chi2-pg1159} 
\end{figure*} 

The results of our period-to-period fit using the \emph{K2}
periods are shown in Fig. \ref{chi2-pg1159}, in which we depict the quality  
function  of the  period  fit in terms of $T_{\rm eff}$  
for the PG~1159 sequences with different stellar masses.   
Black (red) lines correspond to stages before (after) the``evolutionary 
knee'' (see Fig. \ref{logg-teff}). There is a very pronounced minimum of 
the quality function, corresponding to a model of $M_{\star}= 0.565 M_{\odot}$ and $T_{\rm eff}= 129\,613$ K. This model produces the best agreement between observed and theoretical periods. Note that this model is outside the 1$\sigma$ $T_{\rm eff}$ range indicated by the spectroscopy. 
We have also carried out an additional  period fit ignoring the modes with $\ell= 2$, and only fitting the modes $\ell= 1$. 
The result of that period fit indicates the same $\chi^2$ minimum as in the case of the mode fitting with $\ell= 1$ and $\ell = 2$. We conclude that this model constitutes the asteroseismic model for PG~1159$-$035.
This model is very similar to the one derived by \citet{2008A&A...478..869C} considering the \citet{Winget91} and \citet{Costa08} period sets, only differing slightly in temperature. Indeed, the current model 
is $\sim 1600$ K hotter 
than the model derived by \citet{2008A&A...478..869C}. The adopted asteroseismic model corresponds to an evolutionary stage just after the star reaches its
maximum effective temperature (${T_{\rm eff}}^{\rm MAX}= 167\,000$
K; see Fig. \ref{logg-teff}). 

In Table~\ref{tab:per-l1l2m0} we show a  detailed
comparison of the observed periods of PG~1159$-$035 and the theoretical
$m= 0$ periods  of  the  asteroseismic  model. To
quantitatively assess the quality of the period fit, we compute the
average   of   the   absolute   period    differences,
$\overline{\delta \Pi_i}= \left( \sum_{i= 1}^N |\delta \Pi_i|
\right)/N$, where $\delta \Pi_i= (\Pi_{\ell,k}-\Pi_i^{\rm o})$ and $N= 41$,  
and the root-mean-square residual, $\sigma= \sqrt{(\sum_{i= 1}^N
  |\delta \Pi_i|^2)/N}= \sqrt{\chi^2}$.  We obtain $\overline{\delta
  \Pi_i}= 2.12$~s and $\sigma= 3.08$~s.  
To have a global indicator of the
goodness of the period fit that considers the number of free
parameters, the number of fitted periods, and the proximity between
the  theoretical and observed periods, we computed the Bayes
Information Criterion \citep[BIC;][]{2000MNRAS.311..636K}: 
${\rm BIC}= n_{\rm p} \left(\frac{\log N}{N} \right) + \log \sigma^2$,
where $n_{\rm p}$ is the number of free parameters of the
models, and $N$ is the number of observed periods. The smaller the
value of BIC, the better the quality of the fit. In our case, $n_{\rm
  p}= 2$ (stellar mass and effective temperature),   $N= 41$, and
$\sigma= 3.08$\,s.  We obtain ${\rm BIC}= 1.06$, which means that our
period fit is acceptable.  In comparison, \citet{2021A&A...645A.117C} obtain
${\rm BIC}= 0.59$ for the PNNV star RX~J2117+3412, ${\rm BIC}= 1.18$ for 
the hybrid DOV star HS~2324+3944, ${\rm BIC}= 1.15$ for the PNNV star NGC~1501, and ${\rm BIC}= 1.20$ for the PNNV star NGC~2371.
On the other hand, \citet{2022A&A...659A..30C} and \citet{2019ApJ...871...13B} obtain ${\rm BIC}= 1.13$ and ${\rm BIC}= 1.20$, respectively, for the prototypical star of the DBVs class of pulsating WDs, GD~358. The asteroseismic model for PG~1159$-$035 has  
$(\overline{\Delta \Pi_{k}})_{\ell= 1}= 22.02$ s and $(\overline{\Delta \Pi_{k}})_{\ell= 2}= 12.60$ s,
in agreement with the measured values, $\Delta \Pi_1= 21.28$ s  and $\Delta \Pi_2= 13.02$ s. 

We also include in Table~\ref{tab:per-l1l2m0} (column 7) the
rates of period change ($\dot{\Pi}\equiv d\Pi/dt$) predicted for each
$g$ mode of PG~1159$-$035 according to the asteroseismic model. 
Note that all of them are positive
($\dot{\Pi}>0$), implying that the periods in the model are lengthening over
time. The rate of  change of periods in WDs and  pre-WDs is related to the rate of change of temperature with time  $\dot{T}$ ($T$ being the temperature at the region of the period
formation) and $\dot{R_{\star}}$ ($R_{\star}$ being the stellar
radius) through the order-of-magnitude expression 
$(\dot{\Pi}/\Pi) \approx -a\ (\dot{T}/T) + b\ (\dot{R_{\star}}/R_{\star})$
\citep{1983Natur.303..781W}, $a, b$ being positive constants close to 1. 
According  to our asteroseismic
model, the star is entering its cooling stage after reaching its maximum
temperature (Fig. \ref{logg-teff}). As a
consequence, $\dot{T}<0$ and $\dot{R_{\star}} <0$ with $|-a\ (\dot{T}/T)|>|b\ (\dot{R_{\star}}/R_{\star})|$, and then, 
$\dot{\Pi} > 0$. Our  best  fit  model  has all the modes with $\dot{\Pi}>0$, and 
thus it does  not  reproduce  the   
measurements  of  \citet{2008A&A...489.1225C} neither the values shown
in Fig. \ref{Pdot} of the present paper, which  indicate  that  the  
pulsation modes of PG~1159$-$035 have positive and negative values  
of $\dot{\Pi}$
(see Fig. \ref{logg-teff}). 
Also, the magnitude of the observed rates of period change in PG~1159$-$035 are larger than the values derived from the asteroseismic model. This may be because the star 
could have a very thin He envelope, which would inhibit nuclear burning and shorten the evolutionary timescale \citep{2008ApJ...677L..35A}. 
This would result in larger rates of period change. 
We also note that our models do not include radiative levitation, which might 
influence the change in position of the nodes of the eigenfunctions of the pulsation modes, and the photospheric abundances in hot WDs depend on the balance between the flow of matter sinking under gravity and the resistance due to radiative levitation, as well as on the weak residual wind, driven by the metal opacities. It is also possible that the observed period changes are not attributable to stellar evolution, but to another (unknown) mechanism. For instance,
in the case of the DOV star PG~0122+200, the detected rates of period changes are 
much larger than the theoretical models predict as due simply to evolutionary cooling, and it is suggested that the resonant coupling induced within rotational triplets 
could be the mechanism operating there \citep{2011A&A...528A...5V}.


In Table~\ref{table:modelo-sismo}, we list the main
characteristics of the asteroseismic model for PG~1159$-$035. 
The quoted uncertainties in the stellar mass and 
the effective temperature  of the best fit model 
($\sigma_{\rm M_*}$ and $\sigma_{T_{\rm eff}}$) are internal errors resulting 
from the period fit procedure alone, and are assessed according to 
the following expression, derived by \citet{1986ApJ...305..740Z}:

\begin{equation}
\sigma_i^2= \frac{d_i^2}{S-S_0}
\end{equation}

\noindent where $S_0= \chi^2(M_*^0, T_{\rm eff}^0)$ 
is the absolute minimum of $\chi^2$ which is reached at 
($M_*^0, T_{\rm eff}^0$) corresponding to the best-fit model, and 
$S$ the value of $\chi^2$ when we change the parameter $i$ 
(in our case, $M_*$ or $T_{\rm eff}$) by an amount $d_i$ 
keeping fixed the other 
parameter. The quantity $d_i$ can be evaluated as  the minimum  
step in the grid of the parameter $i$. 
We have $d_{T_{\rm eff}} \equiv \Delta T_{\rm eff}\sim  1000$ K
and $d_{M_*} \equiv \Delta M_*$ in the range $0.009-0.024 M_{\odot}$. 
The rest of the uncertainties are calculated based on those in mass and effective temperature.
The effective temperature of the asteroseismic model is lower than the spectroscopic effective temperature $T_{\rm eff}$ of PG~1159$-$035. The
seismic stellar mass ($0.565\pm0.008 M_{\odot}$) is consistent with the 
range of masses indicated by the period spacings of PG~1159$-$035 ($0.54 \lesssim M_{\star}/M_{\odot}\lesssim 0.58$),  and compatible with the spectroscopic mass ($M_{\star}= 0.54 \pm 0.07M_{\odot}$). The luminosity of the asteroseismic model, $\log(L_{\star}/L_{\sun})=  2.19\pm0.04$ is $\sim 20$\%
lower than the luminosity inferred by \citet{2011A&A...531A.146W}, $\log(L_{\star}/L_{\sun})= 2.58$, based on the spectroscopic $T_{\rm eff}$ and the evolutionary tracks of \citet{2006A&A...454..845M}, the same that we use in the present paper. 

\begin{figure}[ht]
\includegraphics[clip,width=1.0\columnwidth]{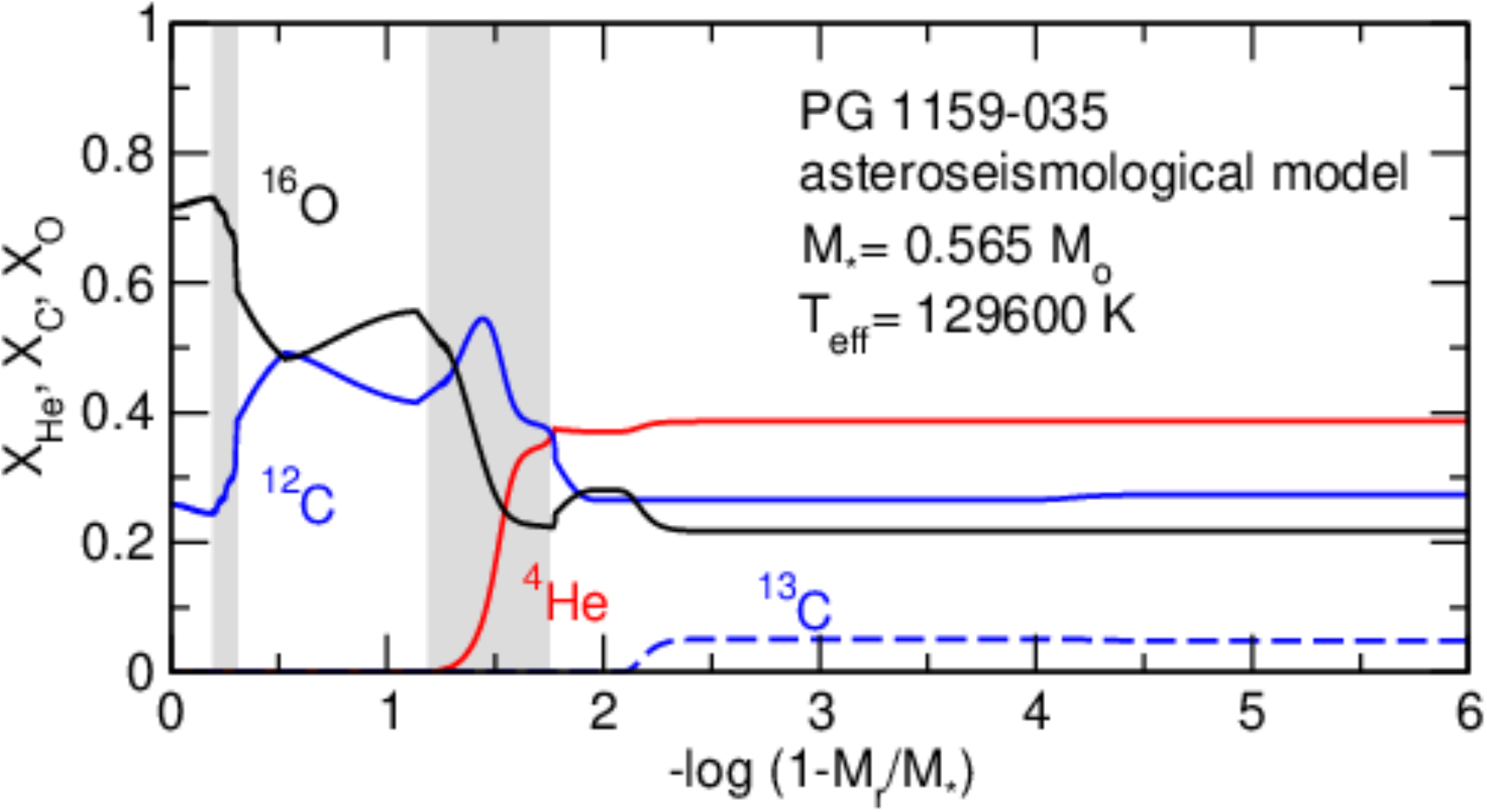}
\caption{Internal  chemical  profile  of  the asteroseismic model of PG~1159$-$035  
($M_{\star}= 0.565 M_{\sun}$, $T_{\rm eff}= 129\,600$ K)  in  terms  of  the  
outer  fractional  mass.  The locations of the O/C and O/C/He chemical interfaces
are indicated with gray regions.}
\label{asteroseismic-model} 
\end{figure}

In Fig. \ref{asteroseismic-model} we display the fractional abundances ($X_i$) of the main chemical species, $^4$He, $^{12}$C, $^{13}$C, and $^{16}$O, corresponding to our best asteroseismic model of PG~1159$-$035,  with $M_{\star}= 0.565 M_{\odot}$  and $T_{\rm eff}= 129\,600$ K. The chemical transition regions of O/C and O/C/He are emphasized with gray bands. The precise location, thickness, and steepness of these chemical transition regions fix the mode-trapping properties of the model \citep[see, e.g.,][for details]{2005A&A...439L..31C,2006A&A...454..863C}. Note that the chemical composition in the models are not free parameters, but the result of the evolutionary calculation. It is therefore not an asteroseismic determination of the envelope composition.

\subsection{Nonadiabatic analysis}
\label{nonadiabatic}

Table~\ref{tab:per-l1l2m0} also provides information
about the pulsational stability/instability  nature of the modes
associated with the periods fitted to the observed ones (eight
column). We examined the sign and magnitude of the computed linear
nonadiabatic growth rates $\eta_k=-\Im(\sigma_k)/\Re(\sigma_k)$, where $\Re(\sigma_k)$ and $\Im(\sigma_k)$ are  the  real  and  the imaginary parts, respectively, of the complex eigenfrequency $\sigma_k$. We have employed the nonadiabatic version of the {\tt LP-PUL} pulsation code \citep{2006A&A...458..259C}, that assumes the ``frozen-in convection'' approximation \citep{1989nos..book.....U}\footnote{We note that this approximation is not relevant in the present case, since PG1159 stars probably do not develop important surface or subsurface convection zones that could impact on g-mode excitation.}.  A positive value of $\eta_k$ means that the mode  is linearly unstable. 

\begin{figure*}[ht]
\includegraphics[clip,width= 1.0\textwidth]{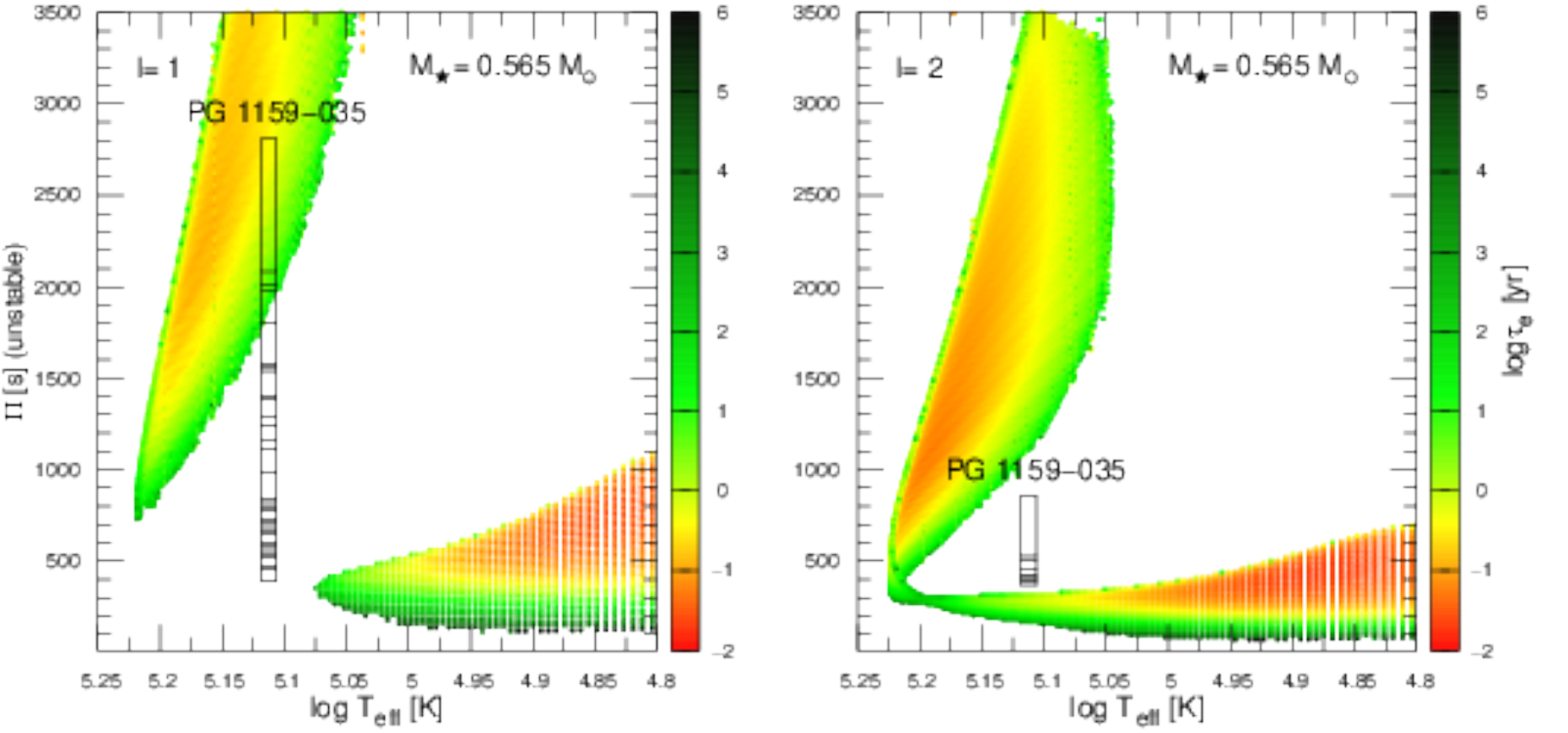}
\caption{Left panel: periods of excited $\ell= 1$ $g$ modes as a function of the effective 
temperature, with the palette of colors (right scale) indicating the 
logarithm of the $e$-folding time ($\tau_{\rm e}$ in years), for  
the PG~1159 sequence with $M_{\star}= 0.565 M_{\odot}$. Right panel: same as left panel, 
but for $\ell= 2$ modes. In both panels, the pulsation periods with the identification of $\ell$ according to our asteroseismic model (see Table \ref{tab:per-l1l2m0}), 
are shown as horizontal segments, where their widths represent 
the possible $T_{\rm eff}$ interval, according to the best asteroseismic model (Table \ref{table:modelo-sismo}).
}
\label{per-l1l2m0} 
\end{figure*} 

We show in Fig.\ \ref{per-l1l2m0} the periods 
of excited $\ell= 1$ (left panel) and $\ell= 2$ (right panel) $g$ modes as a function of the effective 
temperature for the sequence of PG~1159 models with $M_{\star}= 0.565 M_{\odot}$.
In both panels, the identified pulsation periods of PG~1159$-$035 (see Table \ref{tab:per-l1l2m0}), 
are shown as horizontal segments, where the segment length represents the $T_{\rm eff}$ 
range from the best asteroseismic model  (Table \ref{table:modelo-sismo}). For the effective 
temperature and stellar mass of the asteroseismic model, 
all the $\ell= 1$ $g$ modes (left panel) are pulsationally stable, in disagreement
with the existence of $\ell= 1$ excited modes in PG~1159$-$035. 
Excited periods predicted by higher $T_{\rm eff}$ models  
(at the left of the left panel) could explain the long periods shown by the star. 
However, this instability branch corresponds to
models that are before the maximum $T_{\rm eff}$ of the sequence. Our 
non-adiabatic $g$-mode calculations are not able to reproduce the excited  
periods in the star. Regarding the $\ell= 2$ modes (right panel), 
our stability computations predict instability for modes with periods
in the range $68 - 316$ s, thus excluding the interval of quadrupole periods 
excited in PG~1159$-$035 ($365 - 856$ s). We conclude that our 
asteroseismic model, while able to closely reproduce the periods observed 
in PG~1159$-$035, fails to predict their excitation, if the star is after the maximum temperature knee.

\begin{figure*}[ht]
\includegraphics[clip,width= 1.0\textwidth]{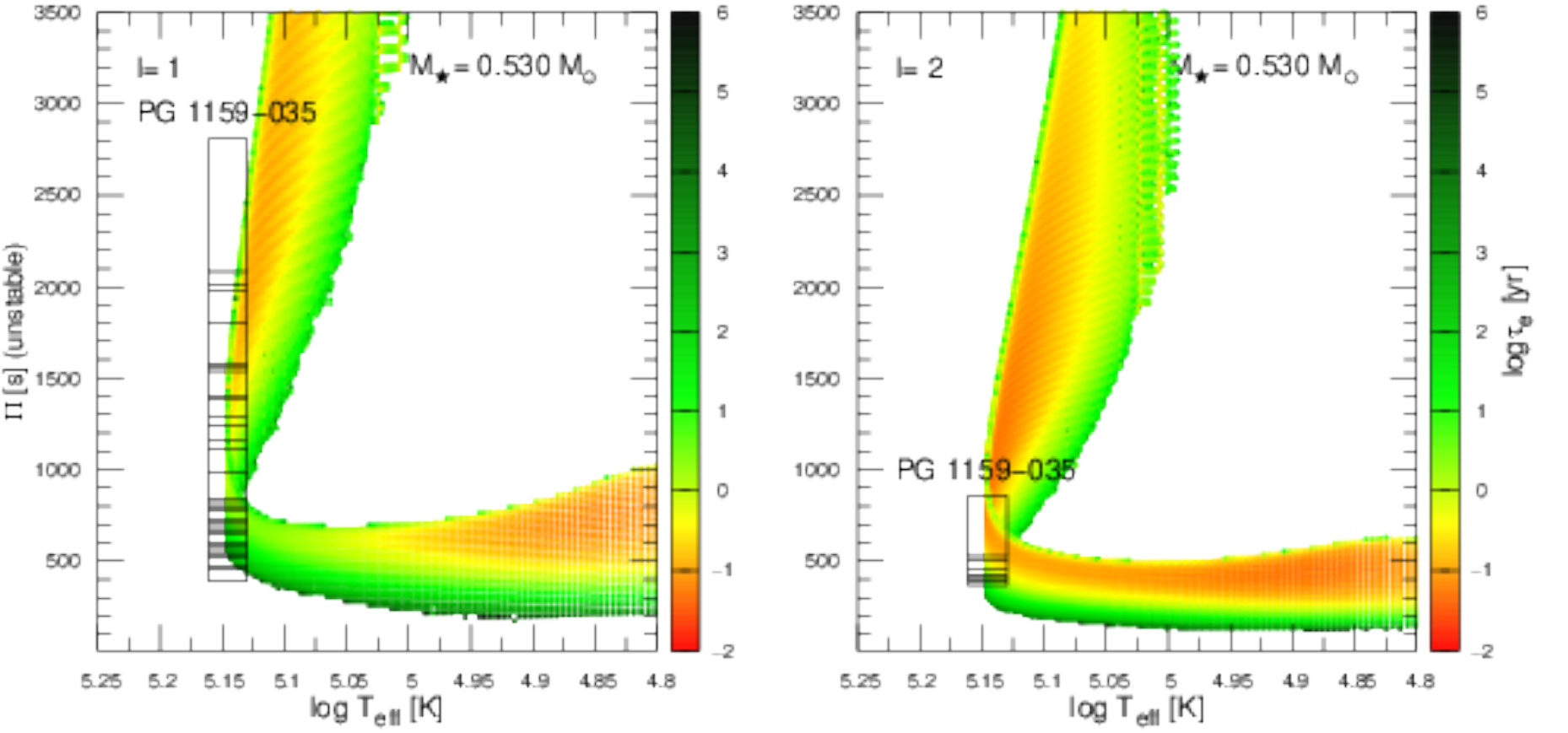}
\caption{Left panel: periods of excited $\ell= 1$ $g$ modes as a function of the effective 
temperature, with the palette of colors (right scale) indicating the 
logarithm of the $e$-folding time ($\tau_{\rm e}$ in years), for  
the PG~1159 sequence with 
$M_{\star}= 0.530 M_{\odot}$. In this case, the effective temperature and its uncertainties 
(horizontal segments) correspond to the spectroscopic determination of 
\citet{2011A&A...531A.146W}. Right panel: same as left panel, 
but for $\ell= 2$ modes. In both panels, the pulsation periods with the identification of $\ell$ according to our asteroseismic model (see Table \ref{tab:per-l1l2m0}), 
are shown as horizontal segments, where their widths represent 
the possible $T_{\rm eff}$ interval, according to spectroscopy
(Table~\ref{table:modelo-sismo}).}
\label{per-l1l2m0-spec0530} 
\end{figure*}

\begin{figure*}[ht]
\includegraphics[clip,width= 1.0\textwidth]{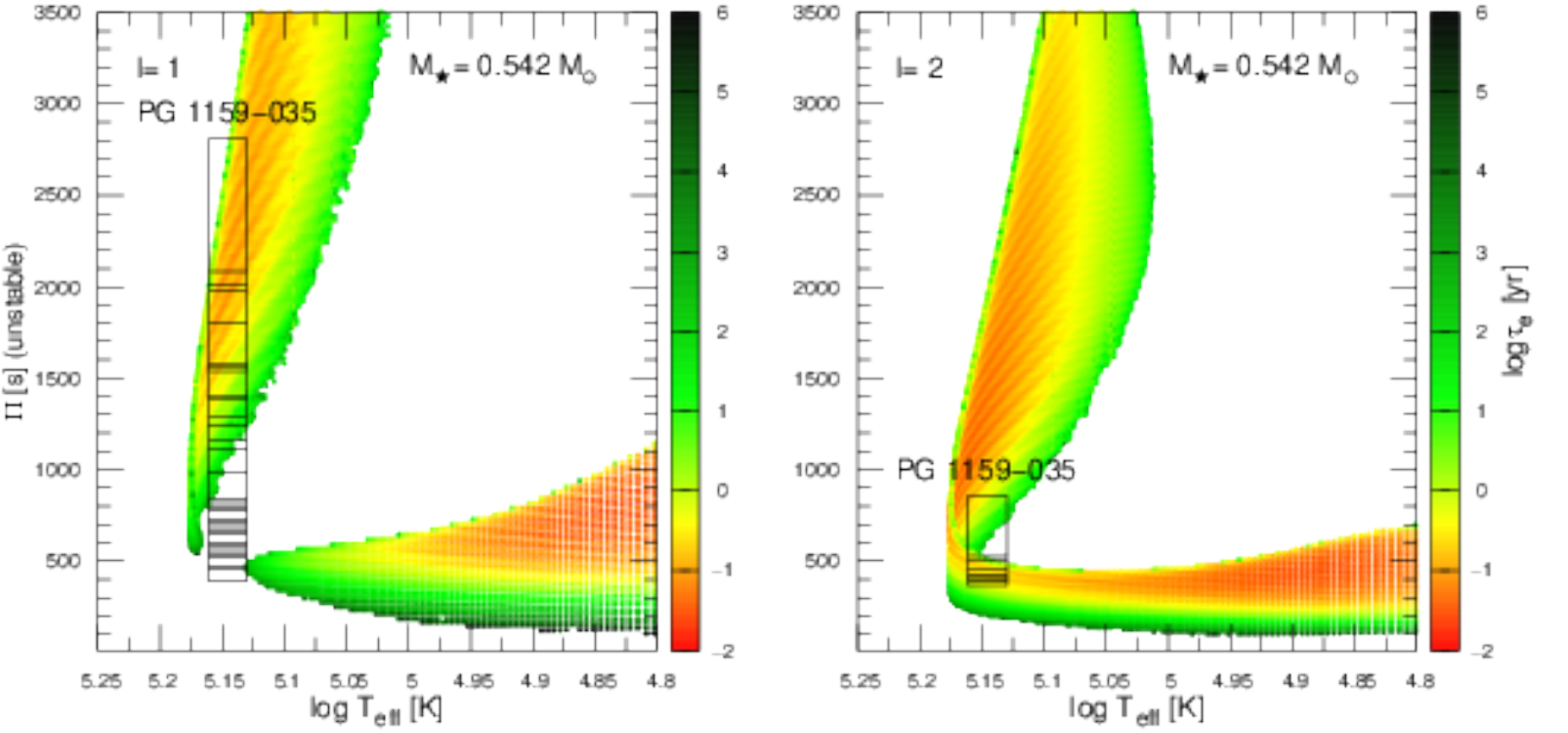}
\caption{Left panel: periods of excited $\ell= 1$ $g$ modes as a function of the effective 
temperature, with the palette of colors (right scale) indicating the 
logarithm of the $e$-folding time ($\tau_{\rm e}$ in years), for  
the PG~1159 sequence with $M_{\star}= 0.542 M_{\odot}$. Right panel: same as left panel, 
but for $\ell= 2$ modes. In both panels, the pulsation periods with the identification of $\ell$ according to our asteroseismic model (see Table \ref{tab:per-l1l2m0}), 
are shown as horizontal segments, where their widths represent 
the possible $T_{\rm eff}$ interval, according to spectroscopy
(Table~\ref{table:modelo-sismo}).}
\label{per-l1l2m0-spec0542} 
\end{figure*}

We expanded our analysis to include the stability of $\ell= 1$ and $\ell= 2$ modes for PG~1159 model sequences with $M_{\star}= 0.530 M_{\odot}$ and $M_{\star}= 0.542 M_{\odot}$ that embrace 
PG~1159-035's spectroscopic mass. The results are displayed in Figs. \ref{per-l1l2m0-spec0530} and \ref{per-l1l2m0-spec0542}. The nonadiabatic calculations for these masses predict unstable $\ell= 1$ modes with periods up to $\sim 1000$ s, and unstable $\ell= 2$ modes with periods up to $\sim 600$ s. However, $\ell=1$ modes with periods longer than $\sim 1000$ s and $\ell=2$ modes with periods longer than $\sim 600$ s are predicted to be pulsationally stable. 

In summary, non-adiabatic calculations considering the best asteroseismic model for PG~1159-305, or adopting stellar models within the range of PG~1159-035's spectroscopic mass are unable to predict the excitation of the long-period $\ell=1$ and $\ell= 2$ modes detected in this star with the data of the \emph{K2} mission.  It is interesting to note that, for GW Vir stars that are evolving at an stage after the "evolutionary knee", current non-adiabatic calculations do not predict the excitation of $g$ modes with periods longer than $\sim 1000$ s \citep{1996ASPC...96..361S, 1997A&A...320..811G,2005A&A...438.1013G,2006A&A...458..259C,2007ApJS..171..219Q}. 
We note, in passing, that these results are robust since in the case of these very hot stars, the pulsational stability analyzes are not affected by typical uncertainties related 
to convection/pulsation interaction, since PG~1159 stars probably do not have significant outer convective zones. In summary, the existence of very long periods in this DOV star is uncertain. A possible explanation is that these modes are the result of nonlinear combination (difference) frequencies (see Sect. \ref{section10}).

We close this section by noting that the precise location of the boundaries of the GW Vir instability domain depends sensitively on the precise value of the abundances of C and O in the driving region of the stars \citep{2004ApJ...610..436Q}.  In particular, by varying moderately the C and O abundances at the driving region, the blue edges of the GW Vir instability strip can be substantially shifted to higher or lower effective temperatures, 
according to the extensive calculations of \citet{2007ApJS..171..219Q}. For instance, if the O abundance changes from 20\% to 40\%, with the C abundance fixed at 40\%, the blue edge of the instability strip for $\ell = 1$ $g$ modes gets hotter by $\sim 10\,000$ K \citep[see Fig. 32 of][]{2007ApJS..171..219Q}. We conclude that a reasonable contrast in the O and C abundances at the driving region of PG~1159$-$035 in relation to the atmospheric abundances, could alleviate the discrepancy between the location of our asteroseismic model and the very existence of pulsations in this star.

\subsection{Asteroseismic distance}
\label{sesimic-distance}

The asteroseismic distance to PG~1159$-$035 can be computed as in \citet{2021A&A...655A..27U}.   Based on the luminosity of the asteroseismic model, $\log(L_{\star}/L_{\odot})= 2.19\pm0.04$,
and a bolometric correction $BC= -7.6\pm 0.2$ from \citet{1994ApJ...427..415K}
\citep[estimated from][]{1991A&A...244..437W}, the absolute magnitude can be assessed as $M_{\rm V}= M_{\rm B}-BC$,  where $M_{\rm  B}= M_{{\rm B},\odot} - 2.5\ \log(L_{\star}/L_{\odot})$.  We employ
the solar bolometric magnitude $M_{\rm B \odot}= 4.74$
\citep{2000asqu.book.....C}. 
The seismic distance $d$  is
derived from the relation: $\log d= [m_{\rm V} - M_{\rm V} +5 -
 A_{\rm V}(d)]/5$. We employ the interstellar extinction law of
\citet{1998A&A...336..137C} for $A_{\rm V}(d)$, which is a nonlinear 
function of the  distance  and  also  depends on  the  Galactic  latitude  ($b$).
For  the  equatorial  coordinates of  PG~1159$-$035  (Epoch  B2000.00, $\alpha=12^h01^m45.97^s$ and  $\delta=-03^d45^\prime 40.62^"$) the corresponding
Galactic latitude is $b = 56.8646$.  We use the
apparent visual magnitude  $m_{\rm V}= 15.04\pm0.01$
\citep{2011MNRAS.410..899F},  and obtain the seismic distance
and parallax  $d= 444^{+69}_{-59}$~pc and  $\pi= 2.25^{+0.35}_{-0.30}$~mas, respectively, using the extinction coefficient  
$A_{\rm V}= 0.064^{+0.002}_{-0.001}$. 
A significant check for the validation of the
asteroseismic model for PG~1159$-$035 is the comparison of the
seismic  distance with the distance derived from astrometry. We
have available the estimates  from {\it Gaia}, $d_{\rm G}= 592 \pm 21$
pc and $\pi_{\rm G}= 1.691 \pm 0.06$ mas.  They are in 
agreement at $1.5\sigma$ level with the asteroseismic derivations, considering the
uncertainties in both determinations, in particular the large
asteroseismic luminosity uncertainty. 

\section{Conclusions}
\label{conclusions}

The amount of asteroseismological information available in a pulsating star is directly proportional to the number of detected independent pulsation modes.  PG~1159-035 is a complex pulsator and rich target for asteroseismic investigation.  We first summarize PG~1159-035's pulsational properties, as revealed in the {\it K2} and {\it TESS} light curves.  Our analysis produced a total of 107 frequencies distributed in 44 separate modes and 9 combination frequencies.  The modes include 32 $\ell=1$ modes and 12 $\ell=2$ modes.  Our investigation of the detected frequencies reveals:

\begin{itemize}
    \item 15 $\ell= 1$ modes consistent with a symmetric $m$ splitting of $\delta\nu = 4.0\pm 0.4\ \mu$Hz.
    \item 9 $\ell = 1$ modes with asymmetric $m$ splitting. These modes show $\delta\nu = 4.08\pm 0.01\ \mu$Hz between $m = 0$ and 1 and $\delta\nu = 2.83\pm 0.06\ \mu$Hz between $m= 0$ and $-1$. The asymmetries are not explained by the presence of a simple magnetic field geometry. We must caution that the asymmetric modes can also be explained by combination frequencies, following \citet{Kurtz15}.
    \item 9 $\ell = 1$ modes with single peaks lacking multiplet structure.
    \item 12 $\ell = 2$ modes with an average splitting of $\delta\nu = 6.8\pm0.2\ \mu$Hz. We note that none of the $\ell = 2$ modes are complete quintuplets.
    \item the identification of a possible surface rotational frequency at $8.904\pm0.003\ \mu$Hz, as well as its harmonic at $17.813\pm0.006\ \mu$Hz, which is roughly 9\,per\,cent faster than the rotation frequency inferred from the $\ell = 1,2$ multiplet splittings.
    \item 9 combination frequencies.
    \item{Several modes with periods between 400 and 1000 seconds show Lorentzian widths consistent with coherence timescale shorter than the observation length.}
    \item{The rates of period change for the highest amplitude modes, separately, do not show a clear pattern and can switch between positive and negative values. But, overlapping, they look to be converging to some value and then scattering again.}
    \item the $\ell= 1$ modes form a sequence with an average period spacing of  $21.28\pm0.02$~s.
    \item the $\ell = 2$ modes form a sequence with an average period spacing of 
    $12.97\pm0.4$~s.

\end{itemize}

\par PG~1159-035 joins the hot DBV PG~0112+104 as the second WD with a photometrically detected surface rotation frequency. The $8.9\ \mu$Hz frequency represents a surface rotation rate of $1.299 \pm 0.002$ days. The frequency splittings of the $\ell=1$ and $\ell=2$ modes indicate a rotation period of $1.4 \pm 0.1$ days. The individual modes sample the rotation in different regions of the star, and we find that the rotational splittings are not constant with the radial node $k$ value. In particular, the high {\it k} $\ell$ = 1 modes that preferentially sample the outer atmosphere show asymmetric splittings. Taken together, PG~1159$-$035's pulsation structure and the surface rotation period provide evidence of nonuniform rotation. PG~1159-035 is an important object for future analysis of the effects of differential rotation and internal structure in a DOV star. 


\par We also present the first detection of combination frequencies in PG~1159-035.  Surface convection is not expected to play a role in this hot object. We find that the fractional temperature changes required to produce the observed pulsation amplitudes are $\approx$ 2.5 times than that of a 12,000~K DAV WD. The second order nonlinearities are correspondingly larger, making the nonlinear response of flux to small temperature changes a plausible mechanism to produce combination frequencies in PG~1159-035.  
The second part of this work focuses on using the detected frequencies to complete a detailed asteroseismic investigation of PG~1159-035.  We summarize the results: 

\begin{itemize}
    \item The average period spacings for $\ell = 1$ and $\ell = 2$ give a mass 
    range of $0.54 - 0.59 M_{\odot}$ consistent with the spectroscopic mass.
    \item The detailed asteroseismic fit includes new high {\it k} modes not included in previous studies. 
    \item The best adiabatic asteroseismic fit model has $T_{\rm eff}=129,600\pm 2\,000$ K, $M_{\star}=0.565\pm 0.008\ M_{\odot}$, $\log g= 7.41\pm0.11$, $\log (L_{\star}/L_{\odot})=2.19\pm0.04$, $\log(R_{\star}/R_{\odot})= -1.61\pm0.05$,
    and $M_{\rm env}=0.017$ $M_{\odot}$. 
    \item The best fit model corresponds to an evolutionary stage just after the star reaches its maximum effective temperature.
    \item The luminosity of the best fit model is consistent at $1.5\sigma$ level with the astrometric parallax from {\it Gaia}.
    \item The rates of period change predicted by the best-fit model are positive for all modes, and thus it do not agree with the observed positive and negative values.
    \item A nonadiabatic analysis considering the best-fit asteroseismic model is unable to predict the excitation of any of the periods detected in PG~1159-035. However, representative models of the star according to its spectroscopic parameters are able to predict the unstable periods, except for the long periods ($\Pi \ga 1000$ s) associated to $\ell$ = 1 modes. We expect that the $g$-mode pulsation periods would be modified if one adopts another model to treat the overshooting in the He-burning stage during the evolution of the WD progenitor star, and that could impact the properties of the seismological model for PG~1159$-$035.
\end{itemize}


\section{Acknowledgments}
This work was partially supported by grants from CNPq (Brazil), CAPES (Brazil), FAPERGS (Brazil), NSF (USA) and NASA (USA).
A.H.C acknowledges support from PICT-2017-0884 grant from ANPCyT, PIP 112-200801-00940 grant from CONICET, and G149 grant from University of La Plata. J.J.H. acknowledges support through TESS Guest Investigator Programs 80NSSC20K0592 and 80NSSC22K0737.
D.E.W. and M.H.M. acknowledge support from the United States Department of Energy under grant DE-SC0010623, the National Science Foundation under grant AST 1707419, and the Wootton Center for Astrophysical Plasma Properties under the United States Department of Energy collaborative agreement DE-NA0003843. M.H.M. acknowledges support from the NASA ADAP program under grant 80NSSC20K0455. 
K.J.B. is supported by the National Science Foundation under Award AST-1903828. GH is grateful for support by the Polish NCN grant 2015/18/A/ST9/00578. SDK acknowledges support through NASA Grant \# NNX16AJ15G, via a subcontract from The SETI Institute (Fergal Mullaly, PI).
This paper includes data collected with the Kepler and TESS missions, obtained from the MAST data archive at the Space Telescope Science Institute (STScI). Funding for the TESS mission is provided by the NASA Explorer Program. STScI is operated by the Association of Universities for Research in Astronomy, Inc., under NASA contract NAS 5–26555. This research made use of Lightkurve, a Python package for Kepler and TESS data analysis (Lightkurve Collaboration et al. 2018). This work has made use of data from the European Space Agency (ESA) mission Gaia (https://www. cosmos.esa.int/gaia), processed by the Gaia Data Processing and Analysis Consortium (DPAC, https://www. cosmos.esa.int/web/gaia/dpac/consortium). Funding for the DPAC has been provided by national institutions, in particular the institutions participating in the Gaia Multilateral Agreement. 
We made extensive use of NASA Astrophysics Data System Bibliographic Service (ADS) and the SIMBAD and VizieR databases, operated at CDS, Strasbourg, France.

\vspace{5mm}
\software{Astropy \citep{astropy13,astropy18}, Lightkurve \citep{lk18}, PERIOD4 \citep{2004IAUS..224..786L}, Pyriod \citep{pyriod}, LPCODE \citep{2005A&A...435..631A}, LP-PUL \citep{2006A&A...454..863C}, TESS-LS (https://github.com/ipelisoli/TESS-LS), TESS-Localize \citep{Higgins22} (https://github.com/Higgins00/TESS-Localize)}.

\appendix

\section{Remaining frequencies}
\par We have subtracted 121 independent frequencies from K2 FT, among these we identified 8 as linear combination, 2 as atmosphere rotation frequency and its harmonic, and we classified 99 as $\ell=1$ or $\ell=2$ modes. The 12 remaining frequencies that we could not classify are listed in the Table~\ref{tab:remain}.

\begin{table}[h]
    \centering
    \begin{tabular}{|ccc|} \hline
         Period [s] & Frequency [$\mu$Hz] & Amplitude [mma]  \\ \hline
        24659.36	&	40.55	&	0.12	 \\ \hline
        21168.43	&	47.24	&	0.14	 \\ \hline
        2345.44	&	426.36	&	0.12	 \\ \hline
        1770.91	&	564.68	&	0.12	 \\ \hline
        1141.07	&	876.37	&	0.12	 \\ \hline
        582.71	&	1716.12	&	0.13	 \\ \hline
        252.19	&	3965.32	&	0.16	 \\ \hline
        226.01	&	4424.51	&	0.26	 \\ \hline
        220.66	&	4531.77	&	0.41	 \\ \hline  
        209.36	&	4776.54	&	0.15	 \\ \hline
        202.89	&	4928.85	&	0.16	 \\ \hline
        200.15	&	4996.25	&	0.12	 \\ \hline
    \end{tabular}
    \caption{Remaining frequencies. The frequencies' uncertainty are on the order of $0.01 \mu$Hz. }
    \label{tab:remain}
\end{table}


\bibliography{main}{}
\bibliographystyle{aasjournal}



\end{document}